\documentclass[12pt,letterpaper]{article}
\usepackage{amsmath,amssymb,amsfonts,amsthm,dsfont,mathtools,lipsum,etoolbox}
\usepackage{booktabs,threeparttable,siunitx}
\usepackage{multirow}
\usepackage{setspace}
\usepackage{color}
\usepackage{endnotes}
\usepackage{enumerate}
\usepackage{float}
\usepackage{makecell}
\usepackage[bottom]{footmisc}
\usepackage{graphicx} 
\usepackage{epstopdf}
\usepackage{tikz}
\usepackage{indentfirst}
\usepackage{latexsym}
\usepackage{lscape}
\usepackage{parskip}
\usepackage{rotating}
\usepackage{caption}
\usepackage[normalem]{ulem}
\usepackage{pgfplots}
\usepackage{bm}
\usepackage{verbatim}
\usepackage{mathdots}
\usepackage{soul}
\usepackage{multirow}
\usepackage{mwe}
\usepackage{arydshln}
\usepackage{xcolor}
\usepackage{subcaption}
\usepackage{titlesec}
\usepackage{placeins}
\usepackage{changepage}
\usepackage[colorlinks=true,allcolors=blue]{hyperref}
\usepackage[left=1.0in,right=1.0in,top=1.0in,bottom=1.0in]{geometry}
\usepackage[shrink=0,letterspace=0]{microtype}
\usepackage[authoryear, round,semicolon,sort&compress]{natbib}

\usepackage{mathptmx}
\usepackage[T1]{fontenc}

\allowdisplaybreaks[4]

\definecolor{mygreen}{rgb}{0.2, 0.7, 0.1}

\makeatletter
\newcommand{\leqnomode}{\tagsleft@true}
\newcommand{\reqnomode}{\tagsleft@false}
\makeatother

\definecolor{lightblue}{rgb}{0.8,0.92,1}
\definecolor{lightorange}{rgb}{1,0.9,0.7}

\definecolor{lightgreen}{RGB}{210,245,210}

\DeclareMathOperator*{\argmax}{argmax}
\newcommand\redsout{\bgroup\markoverwith{\textcolor{red}{\rule[0.5ex]{1pt}{1.4pt}}}\ULon}
\newcommand\bluesout{\bgroup\markoverwith{\textcolor{blue}{\rule[0.5ex]{1pt}{1.4pt}}}\ULon}

\newcommand\T{\rule{0pt}{2.5ex}}

\newcommand{\Var}{\text{Var}}

\newcommand{\Supp}{\text{Support}}
\newcommand{\eps}{\varepsilon}

\makeatletter
\newcommand{\VBig}{\bBigg@{2.7}}
\newcommand{\Vast}{\bBigg@{3.6}}
\newcommand{\VVast}{\bBigg@{5.5}}
\makeatother 

\newcommand{\zerodisplayskips}{%
	\setlength{\abovedisplayskip}{0.15cm}
	\setlength{\belowdisplayskip}{0.15cm}
	\setlength{\abovedisplayshortskip}{0.15cm}
	\setlength{\belowdisplayshortskip}{0.15cm}}
\appto{\normalsize}{\zerodisplayskips}
\appto{\small}{\zerodisplayskips}
\appto{\footnotesize}{\zerodisplayskips}

\titleformat{\section}{\fontsize{12}{14}\selectfont\bfseries}{\thesection}{1em}{}
\titleformat{\subsection}{\fontsize{12}{14}\selectfont\bfseries}{\thesubsection}{1em}{}
\titleformat{\subsubsection}{\fontsize{11}{13}\selectfont\bfseries}{\thesubsubsection}{1em}{}

\titlespacing\section{0pt}{8pt}{3pt}
\titlespacing\subsection{0pt}{6pt}{2pt}
\titlespacing\subsubsection{0pt}{4pt}{2pt}

\titleformat{\section}{\fontsize{14}{15.6}\bfseries\selectfont}{\thesection}{1em}{}
\titleformat{\subsection}{\fontsize{12.5}{14.4}\bfseries\selectfont}{\thesubsection}{1em}{}
\titleformat{\subsubsection}{\fontsize{12.5}{14.4}\bfseries\itshape\selectfont}{\thesubsubsection}{1em}{}

\begin{document}	
	
  	\title{Regulation and Frontier Housing Supply\thanks{
  			We thank Joe Tracy, Anthony Murphy, Peleg Samuels, conference participants at the Conference on Low-Income Housing Supply and Housing Affordability 2022, UEA 2020, ESAM 2021, AFES 2021, ESEM 2024, and seminar participants at Aarhus University, Bar Ilan University, Hebrew University of Jerusalem, Ben-Gurion University of the Negev, Tilburg University, Western Galilee College, The University of Tokyo, Kyoto University, and Hiroshima University for helpful comments and suggestions. We thank Yotam Peterfreund for excellent research assistance.}}
  
  \author{Dan Ben-Moshe\thanks{Department of Economics. Ben-Gurion University of the Negev, dbmster@gmail.com}  
  	\ and David Genesove\thanks{Department of Economics. The Hebrew University of Jerusalem and C.E.P.R., david.genesove@mail.huji.ac.il} 
  }
\date{June 21, 2026}
  
   \maketitle
  \begin{abstract}

  Regulation is a major driver of housing supply, yet often difficult to observe directly. We show that frontier cost, the non-land cost of producing housing absent regulation, is identified from prices and quantities alone, without instruments even when quantity is endogenous in a mean regression. Identification requires regulation to enter as a nonnegative wedge with zero in its conditional support. The difference between price and frontier cost yields the regulatory tax. We apply the approach to new multi-floor, multi-family residential construction in Israel. Accounting for random housing quality, we estimate economies of scale at low heights (minimum efficient scale about five floors), nearly constant marginal cost at middle heights, and an elasticity of substitution between land and non-land inputs of about 0.15 to 0.2 at the greatest heights. The estimated mean regulatory tax is 47\% of housing prices, with substantial variation across locations, and is positively correlated with centrality, density, and prices. We also construct a lower bound allowing quality to differ systematically over location and time, assuming weak complementarity between quality and demand. In 2017, when prices were highest in our sample and the bound is most informative, we bound the mean regulatory tax between 38\% (using a 2km radius) and 53\%.

  \end{abstract}

\noindent \textbf{Keywords:} Housing supply; regulation; regulatory tax; stochastic frontier analysis; real estate

\noindent \textbf{JEL codes:} R31, D24, R52

\newpage

\section{Introduction} \label{se:intro}

Housing economics attributes a major role to regulation in determining housing prices and residential development \citep[e.g.,][]{gyourko2006construction,glaeser2009causes,molloy2020effect}. However, the diverse forms of regulation and their often inconsistent enforcement can make direct observation and quantification difficult \citep[e.g.,][]{gyourko2015regulation,cheung2009incidence}. Our solution is to first estimate frontier cost, the non-land cost of producing housing in the absence of regulation, using only observed prices and quantities, and then estimate the regulatory tax, the money-equivalent extent of regulation, as the wedge between price and this cost. We apply this approach to multi-floor, multi-family housing, using apartment prices per square meter and building height as our price and quantity measures. Such data ordinarily do not identify supply separately from demand without an exclusion restriction or other exogenous variation. The frontier is an exception: when regulation is a nonnegative wedge with zero in its support at each height, frontier cost is identified despite endogenous building height.

Under an initial assumption of homogeneous housing, we show that the lowest observed price identifies frontier average cost (AC) below minimum efficient scale (MES) and frontier marginal cost (MC) above MES. This identifies not only costs at tall-building heights but also the increasing-returns region below minimum efficient scale. Our approach replaces standard identification assumptions of exogenous variation with an assumption on the support of demand and supply shocks, allowing identification despite simultaneity. Supply shocks are taken as the marginal cost differences induced by the regulatory environment. Thus our focus is on building height, the margin through which many regulations, including height limits, floor-area ratios, setbacks, and permitting requirements, are reflected.\footnote{Effective stringency may differ from statutory rules because of exemptions, delays, and discretionary approval. These dimensions are difficult to observe or measure consistently but are reflected in observed building height.} We then account for random housing quality differences using stochastic frontier analysis (SFA) and for systematic differences over location and time with a bounds analysis that relies on (weak) complementarity between quality and demand.\footnote{See \cite{BenMosheGenesoveFrDe} for a more general statement of identification under an `assignment at the frontier' assumption.} A price above frontier cost could instead reflect markups, locational differences in non-land costs, or differences in firm efficiency rather than regulation. Section~\ref{se:ident00} examines each; for example, the frontier is robust to excluding areas where non-land costs plausibly differ, and the estimated tax is not lower in thicker markets, as markups or efficiency differences would imply.

Figure \ref{fig:PQSF} provides intuition for identification of frontier costs for homogeneous housing. Each plotted point represents an observed equilibrium price  and height at the intersection of a supply curve that is shifted up by regulatory constraints and a demand curve. The red curve, tracing the locus of equilibria in unregulated markets as demand increases, is frontier marginal cost above MES (i.e., the firm's inverse supply in the absence of regulation). The blue curve, tracing out the locus of equilibria  with break-even demand as regulation is relaxed, is frontier average cost below MES. For illustrative purposes these curves are drawn as continuous. As the figure suggests, identification of frontier cost, by minimum price at each height, depends on the support of demand and supply shocks, requiring sufficient variation of demand in unregulated markets in the region with diseconomies of scale (i.e., above MES) and sufficient variation in both demand and regulation in the region with economies of scale (i.e., below MES).     

\captionsetup{width=\textwidth,justification=raggedright}
\begin{figure}[tp]
	\hspace*{-2cm}  
	\centering
	\begin{tikzpicture}
		\begin{axis} [
			axis y line=center,  
			axis x line=middle, 
			axis on top=true,
			xmin=0,
			xmax=4.2,
			ymin=0,
			ymax=5,
			height=8cm,
			width=15cm,
			axis lines = middle,
			xtick style={draw=none},xtick={1.71,4.2},xticklabels={${MES}$,Height},
			ytick style={draw=none},ytick={0.85,5},yticklabels={min AC, Price},
			]
						
			\addplot [dashed,thick,black] coordinates {(1.71, 0.0) (1.71, 0.85)} node[circle,fill,inner sep=0.8pt]{};
			\addplot [white] coordinates {(0.65, 0.87) (0.65, 3.0)}  
			node[smooth,black,pos=0.49,left]{$\VVast\{$}  node[black,circle,fill,inner sep=0.8pt]{}
			node[smooth,black,pos=1.0,above]{$(h,p)$};
			
			\addplot [white] coordinates {(0.7, 0.87) (0.7, 3.0)}  
			node[smooth,black,pos=0.76,left]{$\VBig\{$}; 
			\node[label={180:{\footnotesize U}}] at (axis cs:0.32,3.060) {}; 
			\draw[black,->] (axis cs:0.25,3.0) -- (axis cs:0.58,2.6); 
			
			\addplot [white] coordinates {(3.2, 3.02) (3.2, 4.5)}  
			node[smooth,black,pos=0.54,right]{$\Vast\}$}  node[black,circle,fill,inner sep=0.8pt]{} node[smooth,black,pos=1.0,above]{$(h',p')$};

			\node[label={180:{\footnotesize RT=U}}] at (axis cs:4.05,3.24) {};
			\draw[black,->] (axis cs:3.8,3.4) -- (axis cs:3.38,3.8);
			\addplot [domain=0:1.71, dashed] {0.85}  node[black,pos=1,right]{};
 
			\node[label={180:{\footnotesize Frontier inverse supply / MC}}] at (axis cs:4.0,1.24) {};
			\draw[black,->] (axis cs:3.03,1.38) -- (axis cs:2.58,1.8);

			\node[label={180:{\footnotesize Frontier AC}}] at (axis cs:1.54,0.25) {};
			\draw[black,->] (axis cs:1.1,0.37) -- (axis cs:1.3,0.95); 

			\node[label={180:{\footnotesize RT}}] at (axis cs:0.38,0.46) {};
			\draw[black,->] (axis cs:0.25,0.6) -- (axis cs:0.5,1.9);
			
			\addplot [domain=1.71:4.5, smooth, thick,red] {1.1 -x + 0.5*x*x} 
			node[circle,fill,black,inner sep=0.8pt,pos=0.1]{} node[circle,fill,black,inner sep=0.8pt,pos=0.285]{} node[circle,fill,black,inner sep=0.8pt,pos=0.6]{};
			
			\addplot [domain=0.4:1.71, smooth, thick,blue] {(x-1.71)*(x-1.71) + 0.847}  
			node[circle,fill,black,inner sep=0.8pt,pos=0.385]{} node[circle,fill,black,inner sep=0.8pt,pos=0.6]{};
			
			\addplot [domain=1:1.3, smooth, thick,orange] {5.0 -0.6*x}  node[smooth,black,pos=1.0,right]{$D$};
			\addplot [domain=1.1:1.25, smooth, thick,mygreen] {-1.555+5*x}  node[smooth,black,pos=1.0,right]{$S$}	 node[circle,fill,black,inner sep=0.8pt,pos=0.47]{};			
			 
			\addplot [domain=2:2.3, smooth, thick,orange] {7.7 -2.0*x}  node[smooth,black,pos=1.0,right]{$D'$};
			\addplot [domain=2:2.3, smooth, thick,mygreen] {2.1 + 0.6*x}  node[smooth,black,pos=1.0,right]{$S'$} node[circle,fill,black,inner sep=0.8pt,pos=0.51]{};	
			
			\node[circle,fill,inner sep=0.8pt] at (axis cs:1.15, 2) {};\node[circle,fill,inner sep=0.8pt] at (axis cs:1.3, 1.9) {};\node[circle,fill,inner sep=0.8pt] at (axis cs:2.5, 2) {};
			\node[circle,fill,inner sep=0.8pt] at (axis cs:1.7, 4.7) {};\node[circle,fill,inner sep=0.8pt] at (axis cs:1.85, 3.5) {};\node[circle,fill,inner sep=0.8pt] at (axis cs:1.65, 2.5) {};
			\node[circle,fill,inner sep=0.8pt] at (axis cs:1.25, 3) {};\node[circle,fill,inner sep=0.8pt] at (axis cs:1.4, 2.5) {};\node[circle,fill,inner sep=0.8pt] at (axis cs:2.6, 4) {};
			\node[circle,fill,inner sep=0.8pt] at (axis cs:1.35, 3.4) {};\node[circle,fill,inner sep=0.8pt] at (axis cs:1.5, 2) {};\node[circle,fill,inner sep=0.8pt] at (axis cs:2.7, 3) {};
			\node[circle,fill,inner sep=0.8pt] at (axis cs:1.45, 1.6) {};\node[circle,fill,inner sep=0.8pt] at (axis cs:1.6, 3.1) {};\node[circle,fill,inner sep=0.8pt] at (axis cs:2.8, 2.6) {};
			\node[circle,fill,inner sep=0.8pt] at (axis cs:1.55, 3.3) {};\node[circle,fill,inner sep=0.8pt] at (axis cs:1.7, 1) {};\node[circle,fill,inner sep=0.8pt] at (axis cs:2.9, 3.9) {};
			\node[circle,fill,inner sep=0.8pt] at (axis cs:1.65, 3.7) {};\node[circle,fill,inner sep=0.8pt] at (axis cs:1.8, 4) {};\node[circle,fill,inner sep=0.8pt] at (axis cs:3.0, 4.1) {};
			\node[circle,fill,inner sep=0.8pt] at (axis cs:1.75, 1.5) {};\node[circle,fill,inner sep=0.8pt] at (axis cs:1.9, 2.9) {};\node[circle,fill,inner sep=0.8pt] at (axis cs:3.1, 3.5) {};
			\node[circle,fill,inner sep=0.8pt] at (axis cs:1.8, 2.6) {};\node[circle,fill,inner sep=0.8pt] at (axis cs:2.0, 2.8) {};\node[circle,fill,inner sep=0.8pt] at (axis cs:2.2, 4.5) {};
			\node[circle,fill,inner sep=0.8pt] at (axis cs:0.9, 3.4) {};\node[circle,fill,inner sep=0.8pt] at (axis cs:2.1, 2.2) {};\node[circle,fill,inner sep=0.8pt] at (axis cs:1.3, 2.9) {};
			\node[circle,fill,inner sep=0.8pt] at (axis cs:1.0, 2.9) {};\node[circle,fill,inner sep=0.8pt] at (axis cs:2.2, 1.8) {};\node[circle,fill,inner sep=0.8pt] at (axis cs:3.4, 4.2) {};
			\node[circle,fill,inner sep=0.8pt] at (axis cs:1.1, 3.6) {};\node[circle,fill,inner sep=0.8pt] at (axis cs:2.3, 2.9) {};\node[circle,fill,inner sep=0.8pt] at (axis cs:3.5, 4.1) {};
			\node[circle,fill,inner sep=0.8pt] at (axis cs:1.2, 2.6) {};\node[circle,fill,inner sep=0.8pt] at (axis cs:2.4, 2.5) {};\node[circle,fill,inner sep=0.8pt] at (axis cs:3.6, 4.6) {};
			
			\node[circle,fill,inner sep=0.8pt] at (axis cs:1.85, 2.0) {};\node[circle,fill,inner sep=0.8pt] at (axis cs:1.4, 3.5) {};\node[circle,fill,inner sep=0.8pt] at (axis cs:0.6, 4.6) {};
			\node[circle,fill,inner sep=0.8pt] at (axis cs:1.25, 1.7) {};\node[circle,fill,inner sep=0.8pt] at (axis cs:3.02, 2.9) {};\node[circle,fill,inner sep=0.8pt] at (axis cs:0.7, 4.1) {};
			\node[circle,fill,inner sep=0.8pt] at (axis cs:0.9, 2.12) {};\node[circle,fill,inner sep=0.8pt] at (axis cs:2.44, 3.8) {};\node[circle,fill,inner sep=0.8pt] at (axis cs:1.55, 1.4) {};
			\node[circle,fill,inner sep=0.8pt] at (axis cs:1.0, 2.2) {};\node[circle,fill,inner sep=0.8pt] at (axis cs:2.4, 4.34) {};\node[circle,fill,inner sep=0.8pt] at (axis cs:1.9, 1.2) {};
			\node[circle,fill,inner sep=0.8pt] at (axis cs:1.1, 1.5) {};\node[circle,fill,inner sep=0.8pt] at (axis cs:2.65, 3.55) {};\node[circle,fill,inner sep=0.8pt] at (axis cs:2.05, 4.15) {};

		\end{axis}
	\end{tikzpicture}
	\caption{\footnotesize \hspace{-0.3cm} Each point represents an equilibrium price and height. 
	At heights with decreasing economies of scale, 
	the red curve represents the firm's frontier inverse supply.  At heights with increasing economies of scale, the blue curve represents the firm's frontier average cost. The regulatory tax is RT. The deviation from the frontier is $U$.
}
\label{fig:PQSF}
\end{figure}

 The regulatory tax quantifies the impact of regulation in money-equivalent form, as a price-cost wedge, representing both the shadow cost of actually enforced restrictions on construction, as well as delays and additional expenses incurred to circumvent these restrictions.\footnote{The precise definition of regulatory tax is provided in Section \ref{se:devreg}. This wedge is the amount by which regulation raises the effective cost of supply, acting as if the supplier faced this additional cost in an otherwise unregulated market. We take no stand as to whether the regulatory tax is welfare improving or decreasing.} It captures the effective burden of restrictions on vertical development, including formal limits, discretionary approvals, delays, and costly exemptions, rather than any single statutory rule. In an unregulated environment but with the same demand, this tax would induce firms to choose a given building height (i.e., number of floors). Implicitly assuming diseconomies of scale, \cite{glaeser2005manhattan} define the regulatory tax at a given price and height as the price less frontier marginal cost (see Figure \ref{fig:PQSF}). Because of the discreteness of building height, as number of floors, there is a range of prices on the supply frontier at any given height (see Figure  \ref{fig:SFstep}). To address this issue, we amend the definition of regulatory tax to be the maximum of zero and price minus the frontier cost of building an additional floor.
 
The regulatory tax definition needs modifying for heights below MES, where no tax in an unregulated environment would induce firms to build.  To account for such observed buildings, we conceptualize the relevant land areas as covering multiple plots. Then, when demand at minimum average cost falls short of MES, equilibrium absent regulation will consist of some plots developed to MES and others left undeveloped, with average height over all plots equal to quantity demanded. We thus define regulatory tax in the region with economies of scale equal to price less the frontier minimum average cost (see Figures \ref{fig:PQSF} and \ref{fig:ACMC}).

In the ideal scenario of Figure \ref{fig:PQSF} the frontier is identified by the minimum observed price at each building height, while regulation is identified using the given price and the frontier. However, this identification is complicated by unobserved quality, such as additional appliances, flooring quality, underground parking, quality or stage of construction, or exterior aesthetic enhancements. We consider both random quality differences and quality that differs systematically over location and time.

We treat random quality differences as part of measurement errors. These errors, which obscure the frontier, are addressed using SFA methods \citep[e.g.,][]{kumbhakar2022stochastic,greene2008econometric}. In contrast to standard SFA, which typically relies on the skewness of deviations from the frontier and symmetry of errors for identification, our main approach exploits the hierarchical and spatial structure of the data. Assuming that the value of locational amenities varies smoothly over space allows us to posit equal regulatory taxes for nearby buildings of the same height. This enables us to separate regulation from measurement errors by leveraging within-building, between-building, and between-bloc variation in prices, where a bloc is a geographical division used by the Land Registry that averages about 150 apartments in multi-housing-unit buildings.\footnote{In terms of the number of dwellings or households, our geographical unit is on average larger than that of \cite{asher2024residential} for Indian cities, about the same size as that of \cite{ahlfeldt2015economics} for Berlin in 2005, and smaller than that of \cite{harari2024residential} for Brazilian cities. Compared to \cite{bayer2007unified}, our geographical units are smaller than census tracts and about twice as large as census block groups in the San Francisco Bay Area.} Distributional assumptions are then used for likelihood estimation of the frontier, and we report robustness to both symmetric and asymmetric specifications of the deviation from the frontier. Additionally, unlike most SFA models, we allow the frontier and distributional parameters to vary arbitrarily with height. This approach is feasible because the data contain hundreds to tens of thousands of observations at each height, enabling us to perform estimation separately at each height.

Quality may, alternatively, be systematically related to  locational amenities if consumers prefer higher-quality housing in areas with more desirable amenities. In this case, the frontier represents the non-land costs of producing housing with minimal, rather than average, quality. Yet,  without further structure, it is not possible to distinguish the effects of regulation from those of quality above the minimum.  
To address this issue, we additionally assume that, within some distance, structural quality and amenities are weak complements: higher-quality housing is built in locations with at least as desirable amenities.\footnote{The complementarity assumption pertains to new homes only. Theoretically, when housing quality is a normal good, the assumption follows from \citet[eq.~(7)]{brueckner2009gentrification}, that, for newly constructed housing, the increase in bid-rent with income is itself increasing in proximity to the urban area's core, implying higher-income households, and thus higher-quality new construction, are in higher-amenity locations. Empirically, in our setting \cite{genesoveFalk} finds that both mean statistical area salary and price per square meter decline with distance from Tel Aviv; again, if quality is a normal good, then it is likely positively correlated with locational-amenities. Also \cite{helms2003understanding} shows for Chicago that both the incidence of renovation and unconditional mean renovation expenditure (interpretable as housing quality demand) increase with neighborhood median home value and proximity to the center, as well as with other local amenities, conditional on age. More generally, previous work shows that higher income households tend to locate in areas with greater amenities \citep{gaigne2022lives,bayer2007unified}.} This allows us to bound the regulatory tax by comparing frontier costs and prices for nearby buildings.  Similarly, preferences for quality may change over time. Weak complementarity of quality with temporal demand shocks allows us to use construction-year effects in hedonic price regressions for existing homes to identify quality differences across time. 

Our empirical application uses data on newly constructed residential buildings in Israel from 1998 to 2017 and exploits variation in prices across both space and time. This market is characterized by an overwhelming reliance on multi-unit buildings and is particularly suitable for our study because height is a central development margin and enforced regulation varies substantially.\footnote{The Israel Bureau of Statistics reports that of buildings completed in 2022, 83\% contained three or more housing units, and 88\% contained two or more  \citep[Chapter 20, Table 3]{CBS2025}.} Even neighboring buildings may face different levels of enforcement depending on builders' success in securing permits, which they must obtain from at least two different levels of local planning committees, each with considerable discretion \citep[see][]{rubin2019planning,czamanski2011characteristic}.

The estimated frontier exhibits economies of scale at low heights: it falls until MES at five floors, where average cost is about 12\% lower than for a one-floor building, while a mean regression of price on height rises. Marginal cost is then roughly constant over the middle range, about 11 to 24 floors, before rising steeply at the greatest heights. At those heights, substituting non-land inputs for land requires increasingly more non-land input for little saving in land; from 30 to 35 floors, the aggregate arc elasticity is about 0.15 to 0.2. This is among the first estimates of the increasing-returns region of the housing cost function and of the minimum efficient scale.

Turning to regulation, the mean regulatory tax is about 47\% of price, close to the estimates of \cite{glaeser2005manhattan} for Manhattan residential buildings and \cite{cheshire2008office} for UK office buildings, both based on commercial cost data. We reach a comparable figure using only prices and heights. This suggests that suppliers would build taller buildings in unregulated markets, despite the difficulty in building upwards. 
We also find substantial variation in the regulatory tax as a percentage of price, with a standard deviation of about 17\%. We decompose this variation and show that roughly two-thirds is attributable to differences between localities, while the remaining one-third occurs within them. We then characterize the sources of this variation. Between localities, higher regulation is positively correlated with higher average prices and density. Within localities, it is correlated with higher density and proximity to locality centers, but these variables add little explanatory power beyond locality fixed effects. The remaining within-locality variation suggests that regulatory stringency is localized, with at times substantial differences even between nearby buildings. Finally, allowing quality to systematically differ over location and time, we bound mean regulatory tax.  In 2017, when prices were at their peak in our sample---so that the lower bound is especially informative---we estimate a lower bound of 38\% (using a 2km radius) and an upper bound of 53\%.

Estimation of the (mean) housing production function has enjoyed a recent renaissance \citep[e.g.,][]{albouy2018housing,combes2021production,cai2017build,brueckner2017measuring,epple2010new}. 
However, most of this research deals with single family housing, with only a few papers addressing building height. \cite{ahlfeldt2018tall} measure the land price elasticity of height, but disclaim any variation in regulatory conditions in their coverage area.  \cite{henderson2017building} 
focus on uncertain property rights rather than regulation, and take a structural approach.
 \cite{tan2020land} infer the bindingness of observed height restrictions from their effect on the land-price to housing-price relationship, an approach, unlike ours, requiring data on land prices.   

A significant challenge in using housing data, as in many other economic applications, is the difficulty of directly measuring costs and regulations, which are often not fully observable. Hence, quantitative assessment of housing regulation typically infers regulatory effects from the partial correlation of housing market outcomes with observed measures of regulatory strictures, such as the Wharton Residential Land Use Regulatory Index of \cite{gyourko2008new} or the new Wharton index of \cite{krimmel2021wharton}. Early studies were concerned with the capitalization of regulation into mean housing prices \citep[e.g.,][]{katz1987interjurisdictional, pollakowski1990effects}. More recent work has focused on the effect of regulation on housing market response to demand shocks by considering housing price variability \citep{paciorek2013supply}, market supply elasticity \citep{saiz2010geographic}, or income pass-through to prices \citep{hilber2016impact}. 

In contrast, \cite{glaeser2005manhattan} and \cite{cheshire2008office} directly measure the regulatory tax by comparing housing prices to external assessments of construction costs. Our analysis complements that approach but differs in the source of the cost estimates: rather than measuring construction costs externally, we identify and estimate the frontier cost schedule from equilibrium prices and heights. This requires no commercial cost assessments, which are often unavailable, especially in lower-income countries, and may not be comparable across locations and input bundles.\footnote{Even when available, as in commercially available software, the cost estimates may be very detailed and conditional on many finely specified input choices, whose distribution would be needed to aggregate costs to the housing-unit per-square-meter level.}
\cite{krimmel2021tax} compute the regulatory tax as the excess of the intensive value of land, inferred from housing prices, over the extensive value of land, observed from land transactions. However, this method is likely appropriate only for single-family homes. \cite{babalievsky2025} infer the extent of commercial land-use regulation from a spatial general equilibrium model, interpreting the gap between marginal benefit and marginal cost of development as the effective impact of regulation. Our approach is not contingent on any specific equilibrium model.

Measuring housing costs and regulation is important for several policy issues. Building upwards can mitigate urban sprawl by increasing density, offering an alternative to outward expansion \citep[e.g.,][]{nechyba2004urban,brueckner2011sprawl,fu2001site}. Variation in housing regulation across locations may reduce productivity by causing spatial mismatches between labor and capital \citep{hsieh2019housing}, although we are agnostic about the welfare consequences of the regulation we measure here. Additionally, housing deregulation is an important policy tool for checking growing inequality of wealth, particularly if due to increasing land scarcity \citep[e.g.,][]{rognlie2016deciphering}. Understanding the effect of regulation on housing  is crucial for designing effective policies to address these and other related policy issues.

The remainder of the paper is organized as follows.
Section \ref{se:id}  focuses on identification.
Section \ref{se:est} describes the estimators.
Section \ref{se:data} reviews the data.
Section \ref{se:results} presents the empirical results. 

\FloatBarrier
\section{Identification} \label{se:id}

This section presents a demand and supply framework for identifying frontier costs when observing only equilibrium prices and quantities - which, as we will discuss, are essentially heights in our context.
Section \ref{se:supply} analyzes frontier supply and
Section \ref{se:ac} frontier average costs at low heights with economies of scale. 
Section \ref{se:devreg} defines the regulatory tax and
Section \ref{se:bounds} bounds the regulatory tax when allowing for quality to be demand-dependent, over location or time.
Section \ref{se:floorheight}  adjusts prices when households' willingness to pay depends on the apartment's floor and the building's height and Section \ref{se:error} explains how to use SFA techniques to incorporate unobserved random quality as measurement error.
Section \ref{se:ident00} addresses potential critiques by discussing the limitations and assumptions involved in our identification strategy.

\subsection{Frontier Supply (Marginal Cost at Heights Above MES)} \label{se:supply}

This section provides conditions under which frontier supply is identified by the joint distribution of equilibrium prices and {quantities}, in an idealized environment of perfectly competitive markets for a single good produced by equally efficient firms.
Since competitive firms supply only at quantities where there are no economies of scale, this discussion concerns such {quantities} only.
The identifying conditions place no restrictions on the joint distribution of the unobserved and observed variables, 
other than their support. Simultaneity will not be a concern.

Consider multi-floor housing built on parcels of one unit of land each. 
For simplicity, at most one building can be built on each parcel, with the building covering the entire parcel. 
Buildings consist of homogeneous housing units.
Define one unit of housing as a 1-floor building on one unit of land.
Then the quantity of housing in one building is its number of floors.
We observe the price per unit of housing, $p\in (0,\infty)$,
and the number of floors, which we refer to as height, $h\in \{1,2,\ldots\}$,
for each newly constructed building. 

Consider parcel-level supply (analogous to firm supply in basic theory), which includes any regulatory restrictions. 
Since the quantity of housing is the number of floors, a supply curve is a nondecreasing step function that takes only nonnegative integer values. 
It is fully characterized by a sequence of jump discontinuities, $p_1,p_2,\ldots$, where $p_h$ is the marginal cost of the $h$-th floor, i.e., the minimum price at which profit-maximizing suppliers would build $h$ floors under the given regulations; at $p_h$ supply jumps from $(h-1)$ to $h$ floors.
A strict maximum height restriction at $\bar h$ floors would take the form of $p_{\bar h+j} = \infty$ for $j>0$.  More generally, builders may be able to overcome restrictions by sufficient expenditure on legal efforts or lobbying; these additional costs explain the vertical gap between non-frontier (regulated) and frontier (unregulated) supply.\footnote{\cite{cheshire2020trophy} argue that, in London, builders may overcome restrictions by employing `trophy' architects.  The regulatory tax would then include the distortion resulting from the excess outlay on the architect compared to the added value to the buyer. Payments or favors to officials are the more likely tool for overcoming restrictions in our application. }
We derive conditions under which the frontier marginal cost of building the $h$-th floor $p_h^f$ is identified by the minimum price at height $h$.

Next consider, for conceptual purposes only, an \textit{area} with a collection of unit land parcels. 
Consumers consider housing services provided on any parcel as identical to those provided on any other parcel in a given area.\footnote{In using area as a conceptual device, 
one need not imagine a contiguous expanse.
See \cite{piazzesi2020segmented} for evidence of buyers searching over noncontiguous areas.} 
Inverse demand for housing in the area, which is assumed continuous, is therefore a function of the total housing consumed in the area.
Define parcel-level demand as market demand for the area divided by the total number of parcels in the area. 

Figure \ref{fig:SFstep} shows parcel-level supply and demand curves.
The red curve is the inverse frontier  supply curve, the object of interest, while the green curve is some inverse non-frontier supply curve.
The blue curve is inverse demand for a low demand shock, while the orange curve is inverse demand for a high demand shock (violet will be considered later).

Equilibria are at the intersections of inverse demand and inverse supply curves. The figure shows the unique equilibrium for each combination of demand  - low ($D_L$) or high ($D_H$) - and supply  - unregulated ($S_U$) or regulated ($S_R$). The equilibrium with no regulation and low demand is $E_A$. At this equilibrium, price lies between the frontier marginal cost of constructing a 3-floor building, $p_3^f$, and a 4-floor building, $p_4^f$, and so only 3-floor buildings are built.

{The equilibrium with no regulation and high demand is $E_B$}.
At this equilibrium, price equals $p_4^f$ with suppliers indifferent between building 3-floor and 4-floor buildings
{and the market clears at the fraction of 3-floor buildings built.}

The two remaining points show equilibria under supply with regulation.  
The equilibrium with regulation and high demand is $E_C$.
Absent regulation, and at the associated equilibrium price $p_C$, suppliers would  build 4-floor buildings.  
Regulation costs lead suppliers to build only 3-floor buildings.
Similarly, at $E_D$, with low demand, 2-floor buildings are built, although suppliers prefer to build an additional floor.   

\captionsetup{width=\textwidth,justification=raggedright}
\begin{figure}[tp]
		\hspace{-2cm}
	\centering
	\begin{tikzpicture}[scale=1.05]
		\begin{axis} [
			axis y line=center,  
			axis x line=middle, 
			axis on top=true,
			xmin=-0.03,
			xmax=4.8,
			ymin=0,
			ymax=5.4,
			height=8cm,
			width=15cm,
			axis lines = middle,
			xtick style={draw=none},xtick={1,2,3,4,4.8},xticklabels={1,2,3,4,$H$},
			ytick style={draw=none},ytick={0.8,1.6,2,3.5,5.7-0.6*3,5.4},yticklabels={$p_1^f$,$p_2^f$,$p_3^f$,$p_4^f$,$p_C$,$P$},
			]			
			
			\addplot [domain=0:1, thick,dashed,red] {0.8}  node[circle,fill,inner sep=1.5pt,pos=0.00]{} node[circle,fill,inner sep=1.5pt]{};
			\addplot [domain=1:2, thick,dashed,red] {1.6}  node[circle,fill,inner sep=1.5pt]{};
			\addplot [domain=2:3, thick,dashed,red] {2}  node[circle,fill,inner sep=1.5pt]{};
			\addplot [domain=3:4, thick,dashed,red] {3.5}  node[circle,fill,inner sep=1.5pt]{};
			
			\addplot [smooth,thick,red] coordinates {(1, 0.8) (1, 1.6)} node[circle,fill,inner sep=1.5pt]{};
			\addplot [smooth,thick,red] coordinates {(2, 1.6) (2, 2)} node[circle,fill,inner sep=1.5pt]{};
			\addplot [smooth,thick,red] coordinates {(3, 2) (3, 3.5)} node[circle,fill,inner sep=1.5pt]{};
			\addplot [smooth,thick,red] coordinates {(4, 3.5) (4, 4.7)} node[smooth,black,pos=1.0,above]{$S_U$};

			\addplot [domain=0:1, dashed,thick,mygreen] {1.2}  node[circle,fill,inner sep=1.5pt,pos=0.00]{} node[circle,fill,inner sep=1.5pt]{};
			\addplot [domain=1:2, dashed,thick,mygreen] {2.5}  node[circle,fill,inner sep=1.5pt]{};
			\addplot [domain=2:3, dashed,thick,mygreen] {3.5}  node[circle,fill,inner sep=1.5pt]{};
			
			\addplot [smooth,thick,mygreen] coordinates {(1, 1.2) (1, 2.5)} node[circle,fill,inner sep=1.5pt]{};
			\addplot [smooth,thick,mygreen] coordinates {(2, 2.5) (2, 3.5)} node[circle,fill,inner sep=1.5pt]{} ;
			\addplot [smooth,thick,mygreen] coordinates {(3, 3.5) (3, 4.7)} node[smooth,black,pos=0.13,left]{$RT(p_C,3)  \{$} node[smooth,black,pos=1.0,above]{$S_R$};

			\addplot [domain=0.5:4.5, smooth, thick,blue] {4.5 -0.6*x} 
			node[circle,fill,black,inner sep=1.0pt,pos=0.375]{} node[smooth,black,pos=0.34,below]{$E_D$} 
			node[circle,fill,black,inner sep=1.0pt,pos=0.625]{} node[smooth,black,pos=0.6,above]{$E_A$} 
			node[smooth,black,pos=1.0,right]{$D_L$};

			\addplot [domain=0.5:4.5, smooth, thick,orange] {5.6-0.6*x} 
			node[circle,fill,black,inner sep=1.0pt,pos=0.625]{} node[smooth,black,pos=0.6,above]{$E_C$} 
			node[circle,fill,black,inner sep=1.0pt,pos=0.75]{} node[smooth,black,pos=0.73,below]{$E_B$} 
			node[smooth,black,pos=1.0,right]{$D_H$};
		
			\addplot [domain=0.5:4.5, smooth, thick,violet] {3.5-0.6*x};
		\end{axis}
	\end{tikzpicture}
	\caption{\footnotesize \hspace{-0.3cm}  Parcel-level inverse supply and demand curves.}
	\label{fig:SFstep}
\end{figure}

Our empirical analysis conditions on building height.  Consider 3-floor buildings, which are built at $E_A$ (where suppliers want, and are permitted, to build 3-floor buildings),
$E_B$ (where suppliers are indifferent between three and four floors, and some build three floors),
and $E_C$ (where suppliers want to build four floors but permitted only three).
The lowest price among these three equilibria is at $E_A$, which is greater than the minimal price $p_3^f$ required to induce unregulated suppliers to build 3-floor buildings.     

Hence, if the pictured high and low demand curves were the extent of demand variation then $p_3^f$ would not be identified.  
Identification requires a positive probability of frontier supply and a demand curve cutting it at $p_3^f$.
The violet demand curve in Figure \ref{fig:SFstep} is just one such curve that would allow identification.
Note that $E_B$, where the high demand curve intersects the unregulated supply curve, identifies the minimal price to build 4-floor buildings $p_4^f$.
Identification of the frontier supply curve as a whole, then, requires sufficient variation in demand in unregulated markets.

Formally, inverse demand $P^d(h,\eps)$, with random demand shock $\eps$, is assumed continuous in height $h\geq 0$.
Inverse supply is defined by the correspondence $P^s(h,W)=\{p \ | \ p^{W}_h \leq p \leq  p^{W}_{h+1}\}$, with  random supply shock $W$ and $h\in \mathbb{N}$. 
The frontier inverse supply is defined by $P^s(h,f)=\{p \ | \ p^{f}_h \leq p \leq  p^{f}_{h+1}\}$, with  $p^{f}_h =\min\limits_{w\in \Supp(W)}   p^{w}_{h}$, for each $h$. 
An equilibrium $( P, h, \alpha)$ is a price $P \geq 0$, height $h \in \mathbb{N}$, and fraction $0 \leq \alpha < 1$, such that the market clears:  $P= P^d(\alpha(h-1) + (1-\alpha)h,e)\in P^s(h,w)$, 
for some $(e,w)\in \Supp(\eps,W) $. 
Now define  
{\begin{align*}
	P(h)=\{P : (P, h, 0)  \textrm{ or } (P, h+1, \alpha), 0 \leq \alpha <1 \textrm{, is an equilibrium, for some }(e,w)\in \Supp(\eps,W)\}.
\end{align*}
}If there exists $e$ with $(e,f)\in \Supp(\eps,W)$ and $0 \leq \alpha<1$ such that $P^d(\alpha(h-1) + (1-\alpha)h,e) = p^{f}_h$, then $p^{f}_h$ is identified by $\min \{P(h)\}$. In other words, we are assuming sufficient realizations of frontier supply, and demand intersecting it at the frontier price. Note that issues of simultaneity do not arise here.  This identification result suggests the sample minimum price at height $h$ as a natural estimator for $p^{f}_h$.

\FloatBarrier
\subsubsection{Spatial Dependence}

The above discussion considers each building in isolation.  Real estate markets, however, are characterized by spatial dependencies.  The inverse demand curves above can be reinterpreted as residual demands for each building, given prices of other new and existing buildings.  However, the probability statements need further consideration under spatial dependencies.

Spatial dependence here can be either local or global. Locally, price may be sensitive to local density, directly through density's effect on utility and indirectly through price response to local supply.  To account for this,  consider indexing both demand and supply shocks by location, and assume weak dependence so that price dependence diminishes, and approaches independence, as the distance between locations increases \citep[see, e.g.,][]{conley2010spatial}. 
Similarly to the nonspatial framework, for each height we assume a positive probability of a location for which the collective shocks within a neighborhood, outside of which prices are essentially independent, lead to realizations of the frontier supply, and demand intersecting it at the frontier price for that height. 

Globally, spatial dependence may arise from cross-location arbitrage in an at least partially closed market, where prices arise from aggregate supply and local demand.   Then, the price at any location will be determined by an aggregate of market-wide shocks---regulation and demand across all locations, such as the boundary rent curve of \cite{fujita1989urban}---and location-specific shocks---regulation and quality at the location.
Identification will then be ensured if there is a positive probability of no regulation at any given value of locational quality.  The fully encompassing greenbelt example discussed in Section \ref{se:support} would be a case in which identification would fail for some heights, but not others. 

\subsection{Frontier Average Cost at Heights Below MES} \label{se:ac}

In perfectly competitive unregulated markets, firms never construct buildings at heights where there are economies of scale as building at heights at or above MES would always be more profitable. 
However, under regulation, suppliers might build at heights below the frontier's MES.  
Minimum price at such heights could not correspond to frontier supply.  Rather, the minimum price identifies frontier average cost, under conditions shown below.

Figure \ref{fig:ACMC} shows the textbook example of a U-shaped frontier average cost curve, along with its associated marginal cost curve.
For simplicity, we present continuous curves.
The frontier supply function maps prices below minimum AC to height equal zero (i.e., the land is left undeveloped)
and maps prices above the minimum AC to the inverse MC (the red curve in Figure \ref{fig:ACMC}).  
At price equal to minimum AC, suppliers are indifferent between leaving the land undeveloped and building at MES.  
Thus an equilibrium where the parcel-level housing quantity demanded at minimum AC falls short of MES
involves price equal to minimum AC, with some parcels left undeveloped and the remainder developed to height MES, with their shares such that the market clears.\footnote{Consider the general perfectly competitive analysis for identical firms with U-shaped AC curves of mass $N$.  Then the industry supply curve is vertical at zero for price below minimum AC, horizontal from zero to $N \times MES$ at price equal to minimum AC and $N \times MC^{-1}(P)$ for price $P>\min AC$).  
If industry demand intercepts industry supply on the horizontal segment, i.e., $P(0) > \min AC > P(N \times MES)$, for inverse demand $P(\cdot)$, firms are indifferent between producing or not.  Equilibrium entails some firms producing at $MES$ and some not producing.}
An equilibrium where the quantity demanded at minimum AC exceeds MES entails an above minimum AC price and construction on every parcel at a common height above MES.  

Inferring frontier costs at heights below MES thus requires the realization of non-frontier supply.  
Equilibrium $E$ must be generated by some such supply curve intersecting with a demand curve (neither is shown).
However, lower prices at the same height $h$ could also be observed, given appropriate demand and regulated supply shocks. The lowest possible observable price is $p' = AC(h)$, which would be generated by the joint realization of a demand and non-frontier supply that intersect at $E'$.\footnote{Recall that firms are perfectly competitive and that the demand that passes through
	$E$ or $E'$ are market demands scaled down to the parcel, and so firm, level.}
No lower price is possible at $h$; otherwise, firms would suffer losses.

\captionsetup{width=\textwidth,justification=raggedright}
\begin{figure}[tp]
	\hspace{-2cm}
	\centering
	\begin{tikzpicture}[scale=1.05]
		\begin{axis} [
			axis y line=center,  
			axis x line=middle, 
			axis on top=true,
			xmin=0.8,
			xmax=4.9,
			ymin=0,
			ymax=5,
			height=8cm,
			width=15cm,
			axis lines = middle,
			xtick style={draw=none},xtick={2.095,3.151,4.9},xticklabels={$h$,MES,$H$},
			ytick style={draw=none},ytick={2.075,2.5,3,5},yticklabels={min AC,$p'$,$p''$,$P$},
			]
			\addplot [domain=1:3.151, smooth, thick] {x*x - 4*x + 4.75};  
			\addplot [domain=3.151:5, smooth, thick,red] {x*x - 4*x + 4.75} 
			node[smooth,black,pos=0.37,right]{$MC$};
			\addplot [domain=1:3.151, smooth, thick,blue] {x*x/3 - 2*x + 4.75+1/x};
			\addplot [domain=3.151:4.6, smooth, thick] {x*x/3 - 2*x + 4.75+1/x} 
			node[smooth,black,pos=1,right]{$AC$};
			\addplot [domain=0:3.151, dashed] {2.075}  node[black,pos=1,right]{};
			\addplot [domain=0:2.095, dashed] {2.5}  node[black,pos=1,right]{};
			\addplot [domain=0:2.095, dashed] {3}  node[black,pos=1,right]{};
			\addplot [dashed] coordinates {(2.095, -1) (2.095, 3)} 
			node[smooth,black,pos=0.87,left]{$RT(p'',h) \quad $}
			node[smooth,black,pos=0.88,left]{$\Bigg\{$}
			node[smooth,black,pos=0.9,right]{$E'$}
			node[smooth,black,pos=1.01,right]{$E''$};
			\addplot [dashed] coordinates {(3.151, -1) (3.151, 2.075)}; 
			\node[circle,fill,inner sep=1.5pt] at (axis cs:2.095, 2.5) {};
			\node[circle,fill,inner sep=1.5pt] at (axis cs:2.095, 3) {};
		\end{axis}
	\end{tikzpicture}
	\caption{\footnotesize \hspace{-0.3cm} {Frontier AC and MC curves.}}
	\label{fig:ACMC}
\end{figure}

Hence, whereas minimum price, conditional on height, converges to MC at heights for which AC is increasing, it converges to AC where AC is decreasing.  Minimum price thus identifies the maximum of frontier AC and MC, denoted as $G(h) = \max\{AC(h),MC(h)\}$, which in Figure \ref{fig:ACMC} is the blue curve $\min \{P(h)\}$ $=AC(h)$ and the red curve $\min \{P(h)\}=MC(h)$.
Whereas identification at heights of increasing AC requires variation in demand in unregulated markets, identification at heights of decreasing AC requires variation in both demand and regulation.

Assuming a U-shaped frontier average cost curve is an important simplification.  
In principle, the cost structure might differ. 
First, average costs might be declining for some region at high heights. 
However, the maximum extent of the rate of decline decreases with height, since total costs are weakly increasing
$
(AC(h)-AC(h-1))/AC(h-1) \geq -1/h.
$
Second, there may be regions where marginal frontier costs exceed average costs yet are decreasing, where firms would ordinarily not operate, but might under regulation.
This would be especially difficult to handle as the minimum observable price would actually exceed frontier marginal costs. 
Furthermore, incorporating such irregular cost structures would involve multiple local turning points,
as opposed to the single one at MES that we have here. For these reasons,  we impose the condition of a U-shaped average cost curve.

\FloatBarrier
\subsection{Regulatory Tax} \label{se:devreg}

Define the regulatory tax as:
\begin{align}
	RT(P,h):=
	\begin{cases}
		P - AC(MES), & h<MES,\\
		\max\{0,\, P - MC(h+1)\}, & h \geq MES,
	\end{cases}
	\label{RT}
\end{align}
where $MES=\arg\min_{h\in\mathbb{N}} AC(h)$. This is the minimum tax in an unregulated environment for which one would observe a height $h$ with price $P$. Define the regulatory tax rate (RTR) as $RT(P,h)/P$.  Both $RT(P,h)$ and $RTR(P,h)$ summarize a local price-cost wedge at height $h$.

Buildings with heights below MES cannot be rationalized in a perfectly competitive market.  We thus define the regulatory tax as the amount needed to rationalize an equivalent average per-parcel quantity over an \textit{area}, as defined in subsection \ref{se:supply}, encompassing the parcel. 
The only possible equilibrium price in an unregulated market consistent with an average quantity less than MES is a price equal to minimum average cost $AC(MES)$ plus the regulatory tax.  In such an equilibrium, parcel-level height demanded is $h$ and firms are indifferent between not building at all and building to MES. Some parcels are left undeveloped and others built to MES, with the share such that demand equals supply.
Hence, at $E$ in Figure \ref{fig:ACMC}, the regulatory tax is
$RT(p'',h)=p''-AC(MES)$.

Above MES, for an unregulated competitive firm to choose height $h$, we must have $MC(h) \leq 	p  \leq  	MC(h+1)$.
Thus when price is below the marginal cost of adding another floor, the regulatory tax is zero
and when price exceeds the marginal cost of adding another floor, the regulatory tax is equal to the difference.
Hence, at $E_C$ in Figure \ref{fig:SFstep},  the regulatory tax is 
$RT(p_C,3)=	\max\{0,p_C-MC(4)\}=p_C-p^f_4$, which would raise marginal costs so that 3-floor buildings would be built absent other regulation.

\subsection{Bounds for Systematic Quality} \label{se:bounds}

 Unregulated suppliers will build better when building higher if households with greater willingness to pay for  locational amenities also prefer higher quality housing, or if households prefer higher quality when purchasing in better locations, or in periods with greater demand for housing. This will result in quality differing systematically over location and time. 
Consequently, the frontier will represent the non-land costs of producing minimal-quality, rather than average-quality, housing,  but the difference between the price and frontier will be the sum of regulatory effects and the excess of quality above the minimum-quality frontier, requiring some method to separate the two.\footnote{Independent measurement error and random quality are incorporated in Section \ref{se:error}; the discussion here concerns only the systematic component of quality that varies with location, or period.}  In this section, we bound the regulatory tax. 

To begin, assume total costs are $C(h)+zh$, where $C(h)$ is the frontier-quality cost of building to height $h$ and $zh$ is the extra cost of building at quality $z \geq 0$; with this specification, additional quality adds the same amount to marginal as to average cost, and profit-maximizing quality is independent of height and thus of regulation. 
Now, for any building $i$, its price $P_i$ is the sum of frontier cost $G(h_i)=\max \{MC(h_i),AC(h_i)\}$, marginal cost due to quality $z_i$, and deviation $U_i\geq 0$,
\begin{align}
	P_i=G(h_i)+z_i+U_i   .\label{eq:decomp0}
\end{align} 
The deviation captures the regulatory effects, and when cost curves are continuous, and suppliers build above MES, as in Figures \ref{fig:PQSF} and \ref{fig:ACMC}, then the deviation is exactly the regulatory tax.

Since $z_i \geq 0$, an upper bound for the regulatory tax is obtained when $z_i=0$,
\begin{align*}
	 U_i  &\leq P_i - G(h_i) ,\\
		RT_i :=RT(P_i,h_i)
	&\leq \bigg\{\begin{array}{ll}
		P_i -	AC(MES), & h_i<MES,\\
		\max\{0,P_i- 	MC(h_i+1)\}, & h_i \geq MES,
	\end{array} 
\end{align*}
This bound, the same as \eqref{RT}, assigns the entire difference between price and frontier to regulatory restrictions, dismissing any contribution from quality.

Next, for a lower bound on the deviation for focal building $i$, consider a comparison building $j$. Taking the difference between equation \eqref{eq:decomp0} for buildings $i$ and $j$, rearranging, and using the nonnegativity of the deviation for building $j$, $U_j \ge 0$, yields a bound for the focal building's deviation: 
\begin{align}
 U_i \geq  \underbrace{(P_i-P_j)}_{(i)}-\underbrace{(G(h_i)-G(h_j))}_{(ii)}-\underbrace{(z_i-z_j)}_{(iii)}. \label{eq:dev0}
\end{align}
Of these three components, we now focus on the quality differential (iii), as it is not observed, and must be inferred through additional structure.  	 
To that end, decompose $z_i-z_j = (z_i - z(a_j,t_i)) + (z(a_j,t_i)-z_j)$. Here, $z(a_j,t_i)$ represents the quality that would arise at the comparison building's location but at the focal building's transaction period. The spatial component, $(z_i - z(a_j,t_i))$, represents the quality difference due to different locations, at the focal building's transaction period. The temporal component, $(z(a_j,t_i)-z_j)$, represents the quality difference due to different transaction periods, at the comparison building's location.

To bound the spatial component, write the price of housing with amenities $a$, transaction time $t$, and quality $z$ as $P(a,z,t)$.
We assume local (weak) complementarity between amenities and quality, i.e., the returns to  quality are nondecreasing with amenities: $P_{za}\ge 0$.\footnote{We use the standard notation $f_{x}$ to denote the partial derivative $\partial f/ \partial x $.}  
This still allows for different trade-offs between amenities and quality in different geographic areas; indeed, imposing global complementarity between amenities and quality would be inconsistent with a constant quality frontier.

A profit-maximizing, price-taking supplier, unconstrained in choice of quality, will choose quality $z(a,t)$ to satisfy the first order condition
\begin{align}
	P_z(a,z(a,t),t) = 1. \label{eq:focz}
\end{align}
For the spatial component, fix time $t$.
Totally differentiating the first order condition \eqref{eq:focz} and price $P(a,z,t)$ implies,\footnote{We solve $dP=P_ada + P_zdz $ and $0=P_{za}da + P_{zz}dz$ for  unknown $dz$ and $da$. } 
\begin{align}
	dz = \frac{1}{1-(P_{zz}P_a/P_{za})} \times dP \equiv \kappa_S(a,z) \times dP. \label{eq:kappa}
\end{align}
Weak complementarity $P_{az}\geq 0$, the second order condition $P_{zz}\leq 0$, and $P_{a}>0$ (by definition) imply $0\leq \kappa_S(a,z) \leq 1$.  Thus if a building's locational amenity is smooth in location, we can conclude that $z(a_j,t_i) - z_i \approx \kappa_{Si} \times (T_{ij}P_j - P_i)$ for all comparison buildings $j$ sufficiently close to focal building $i$, and for some $\kappa_{Si}\in [0,1]$, where $T_{ij}P_j$ is defined as building $j$'s price deflated to building $i$'s transaction period using a housing price index. 

Were systematic quality to vary only spatially and not temporally, the focal building deviation would thus be bound from below by fraction $1- \kappa_{Si}$ of the excess of that building's price over a neighboring building's time-adjusted price, less the difference in the frontier costs at their respective heights - and thus by the maximum of this for each comparison building. As $\kappa_{Si}$ is unknown, we could then obtain a lower bound by choosing the $\kappa_{Si}\in [0,1]$ that minimizes this maximum lower bound. 

However, we need also account for the temporal component. Assume, for newly constructed housing, the standard hedonic price specification  $ P(a,z,t) = \exp(\gamma(t))P^0(a,z) $, so that $\gamma(t)$ are time fixed effects in the log-linear specification. 
Importantly, this builds in complementarity, as $P_{\gamma(t)z} = P_{z} \geq 0$.
Fix amenity $a$.
Totally differentiating the log first order condition for quality \eqref{eq:focz}  and log price, we obtain\footnote{We solve $d\ln P = d\gamma (t) + (P_z/P)dz$ and  $0=d\gamma (t) + (P_{zz}/P_z)dz$ for unknown $dz$ and $d\gamma (t)$.}
\begin{align}
	dz =  \frac{\delta(a,z)}{ 1+\delta(a,z)} \times dP \equiv \kappa_T \times dP,	\label{eq:kappa1}
\end{align}
using the first order condition $P_z=1$, and where $\delta  \equiv -P_z^2/(PP_{zz}) \ge 0$ is an inverse measure of the convexity of $P$ as a function of $z$ (a constant for $P$ isoelastic in $z$).  This allows us to write $z(a_j,t_i) - z_j \approx \kappa_T \times (T_{ij}P_j - P_j)$ for all comparison buildings $j$ sufficiently close in time to focal building $i$. 

In contrast to the coefficients $\kappa_{Si}$ for the spatial component, $\kappa_T$ can be estimated.  
Generalizing our price specification above to accommodate existing homes, and noting that the choice of quality for housing constructed at time $t$ can be written as $z(a,\gamma(t))$, let the log price of housing constructed in period $s$ and sold in period $t$ be $\ln P = \gamma(t) + \ln P^0(a,z(a,\gamma(s)))$. 
Then a linear approximation of the price around the quality of new construction at an arbitrary time period 0, $z(a,\gamma(0))$, is\footnote{This follows from $\frac{\partial P^0}{\partial \gamma(s)} = \frac{P_z}{P} \cdot \frac{\partial z}{\partial \gamma(s)} =\frac{P_z}{P} \cdot (-\frac{P_z}{P_{zz}}) \equiv \delta$.}
\begin{align}
	\ln P \approx  \gamma(t) + \ln P^0(a,z(a,\gamma(0)))+ \delta(a,z(a,\gamma(0))) \cdot \gamma(s). \label{eq:RT_i}
\end{align}

This motivates estimating $\delta$ by the proportionality coefficient in a restricted log price regression that conditions on the dates of transaction (`period effect') and construction (`cohort effect'), with the cohort effect constrained to be proportional to the period effect, and with parcel fixed effects for $\ln P^0(a,z(a,\gamma(0)))$.\footnote{In principle, $\delta$ can vary across locations.  However, allowing $\delta$ to vary by locality in the empirical analysis does not change our results.  That issue, along with depreciation and the relationship of the proportionality restriction to the well known period-cohort-age  problem are discussed further in  Appendix \ref{ap:timeeffects}.}

Returning to inequality \eqref{eq:dev0}, inserting the approximations for the spatial and temporal components of the quality differentiation, accounting for discrete height and nonnegativity of the focal building's own deviation, and noting that the inequality holds for all local buildings, which includes the focal building itself, we choose the largest bound for the set $\Omega_i(d)$ of buildings $j$ within a radius $d$ from building $i$. The  lower bound is now obtained by a minimax, 
{\fontsize{11.35}{14.0}
	\begin{align}
				&		 U_i \gtrapprox \min_{\kappa_{Si} \in [0,1]} \max_{j \in \Omega_i(d)}  \{ [G(h_j) - G(h_i)] - [(P_j-P_i) -\kappa_T (P_j-T_{ij}P_j) - \kappa_{Si} (T_{ij}P_j-P_i) ]\} , \label{eq:devbound} \\
		&		{RT}_i \gtrapprox \min_{\kappa_{Si} \in [0,1]} \max_{j \in \Omega_i(d)} \max \{0, [G(h_j) - G(h_i+1)] - [(P_j-P_i) -\kappa_T (P_j-T_{ij}P_j) - \kappa_{Si} (T_{ij}P_j-P_i) ]\} . \label{eq:RTbound}
\end{align}
}Thus, the regulatory tax is bounded from below by the difference between the frontier-quality construction costs of any sufficiently close building $j$ and those of the focal building, minus the difference in their quality-adjusted prices. Choosing the radius $d$ involves a tradeoff: a larger $d$ results in higher lower bounds but reduces the accuracy of the spatial component in the quality approximation. Therefore, we consider how the lower bound changes with respect to $d$.

\subsection{Adjusting Prices for Consumer Preferences of Apartment Floor and Building Height} \label{se:floorheight}

We account for consumers valuing apartment floor or building height by ``efficiency unit" modeling of housing services, 
with log price 
\begin{align}
	\ln (\text{price}) &=  \ln p + \ln m(f, h), \label{Price}
\end{align}
where $m$ is an unknown function representing the premium that all households are assumed willing to pay for an $f$th-floor apartment in an $h$-floor building, and $p$ is the price net of this, reflecting the value of the building's location. 
Hence, per unit of land the quantity of housing in an $h$-floor building is the sum of the premiums, $q(h)=\sum_{f=1}^{h} m(f, h)$.

Although building height maps one-to-one to the quantity of housing (and in our data they are very close, with $0.05 \leq (q(h)-h)/h \leq 0.1 $), they are not identical.
Since the discrete levels of quantity will not be integers, it will usually be convenient to express cost as a function of height.  
Yet, with price stated per unit quantity, we make this relationship explicit. 
Let $h(q)$ denote the inverse of $q(h)$.\footnote{This inverse exists as long as $m(f,h)>0$, for all $1\leq f\leq h$, which is the case empirically.}  
Then $\widetilde C(q)= C(h(q))$, where $\widetilde C(q)$ is the frontier cost of building quantity $q$ and $C(h)$ the frontier cost of building to height $h$.  

Break-even market price for an $h$-floor building  is
\begin{align*}
	AC(h)&=  \frac{C(h)}{\sum_{f=1}^{h} m(f, h)}=\frac{\widetilde C(q(h))}{q(h)}. 
\end{align*}
This is the lowest possible observed adjusted price in a region with economies of scale.  

For diseconomies of scale, the lowest possible observed adjusted price at any given height equals the marginal cost savings from building the next lowest feasible quantity, 
\begin{align*}
	MC(h)&= \frac{C(h)- C(h-1)}{\sum_{f=1}^{h} m(f, h)-\sum_{f=1}^{h-1} m(f, h-1)}= \frac{\widetilde C(q(h))-\widetilde C(q(h-1))}{q(h)-q(h-1)}
	.
\end{align*}

\subsection{Measurement Error and Random Quality} \label{se:error}

Our empirical analysis also allows for random quality differences at both the building and apartment levels.\footnote{For a demand and cost specification that includes both systematic quality $z_s$ and random quality $z_r$, specify price as $P(a,z_s) + z_r$ and cost as $C_0(h) + h \times (z_s + z_r)$, so that firms will be indifferent over choices of random quality. \label{fn:1}} 
While it is not necessary to classify systematic and random types of quality differences, evident sources of random quality differences include varying stages of construction completion at time of transaction \citep[as noted in][]{combes2021production} and small capital goods, such as appliances, that are available at the same cost to both suppliers and buyers, and are part of the apartment price.  We treat such quality differences as measurement errors and apply techniques from SFA.\footnote{Unlike SFA, which assumes unregulated markets with deviations representing firm inefficiency, our approach assumes equally efficient firms, with deviations representing regulation.  Allowing for a region of increasing returns to scale, as in Section \ref{se:ac}, also differs from SFA.}   Literal measurement errors---such as transcription mistakes or misreports of apartment price or floor area---are also considered part of these errors.

Separating the convolution of deviations and errors without additional information can be achieved by restricting their distributions \citep[e.g.,][]{schwarz2010consistent, florens2020estimation}. In practice, SFA often identifies deviations by imposing skewness on deviations and symmetry in measurement errors. We prefer not to rely on shape restrictions such as symmetry and instead leverage the hierarchical structure of the data \citep{kotlarski1967characterizing},\footnote{Our estimates using the hierarchical structure indicate relatively symmetric deviations.} assuming constant-quality prices vary smoothly over space. This then implies equal deviations for sufficiently nearby buildings \textit{of the same height}. Variances are now identified using variation in prices within buildings, across buildings, and across blocs.  Additionally, estimation conducted separately at each height allows these variances, the frontier, and the distributional parameters to vary arbitrarily with height.

\subsection{Threats to Identification and Interpretation} \label{se:ident00}

Identification of the frontier and the regulatory tax only requires observable prices and quantities (heights).
There is no need for exogenous variation or for parametric or separable restrictions on demand or on (regulated or unregulated) supply.\footnote{As discussed in Section~\ref{se:error}, the distributions of deviations from the frontier must be allowed to depend on height. This is formalized in the estimation framework in Section~\ref{se:est_model}, where the frontier and the distributional parameters are indexed by height $h$.} 
Other characteristics of the environment become critical, though.

To be explicit, the regulatory tax is defined as the gap between the transaction price and the frontier at a given height. For this gap to be interpreted as regulatory burden, we must rule out other sources of price--frontier differences, most importantly output-market markups and differences in non-land costs or firm inefficiency. In addition, interpreting the frontier as unregulated non-land costs relies on the support condition that regulation is nonnegative and that zero regulation is attainable (in the sense that the support includes values arbitrarily close to zero). Finally, it requires ruling out below-cost pricing or government subsidization of construction costs.

\subsubsection{Price Taking in the Output Market} \label{se:pricetaking}

As in the empirical housing production function literature, we assume that firms are price takers in the output market \citep[e.g.,][]{albouy2018housing,combes2021production,cai2017build,brueckner2017measuring,epple2010new}. This assumption matters for identification if firms have market power and markups vary across firms, locations, or heights, as then price--frontier gaps may reflect markups rather than regulatory burden.

Our setting suggests that price taking is a reasonable approximation. Our data consist of apartments in generally urban environments characterized by high density and multi-family housing. The new buildings in our sample are, on average, located within a 500 meter radius of an existing population of 5,260 people, or about 1,600 apartments. This suggests thick local markets, with firms facing competition not only from other new construction but also from the existing housing stock, both renovated and unrenovated. Consistent with this, the annual construction flow is about two percent nationally, so the stock of existing homes is much larger than the flow of newly constructed housing. 

A related concern is that large firms may acquire multiple adjacent parcels and therefore control a substantial share of new supply in a small area. Our transaction data do not identify the builder, so we cannot directly measure seller concentration in the output market. We therefore provide indirect evidence on competition in development activity. Nationally, the Israeli construction industry is structurally competitive, with a ten-firm concentration ratio of 0.15 only \citep{MinFin2017}. Local concentration is also likely to be low: the larger firms operate throughout the country,\footnote{The country is about the size of New Jersey, with about half its area a semi-arid, lightly populated desert.} and the locality-level Herfindahl concentration index of auctioned-off building rights for housing units on government-owned land is 0.025, equivalent to forty equally sized firms (Appendix \ref{ap:comp}). While these auctions pertain to competition for construction rights (an input-side margin) rather than to pricing of completed units, they suggest that local concentration in development activity is limited. 

Thus, the structural conditions in the market suggest that markups should not be a major problem here, but it is instructive to consider what types of markups would threaten identification. Write the transaction price as the sum of frontier cost, the regulatory tax, and a markup. A constant absolute markup among all buildings (i.e., constant across firms, locations, and heights) would be incorporated into the measured frontier, simply shifting it up by a constant. Therefore, a constant markup would not bias the inferred regulatory tax, which is identified as the gap relative to the shifted frontier. The concern is instead heterogeneity in markups: if markups vary across firms, locations, or heights, then the inferred regulatory tax will generally reflect the sum of the regulatory tax and the markup in excess of the minimum markup among unregulated buildings at that height.

Whether the estimated regulatory tax is smaller in thicker markets is informative about this potential bias. Since markups should decrease with market size \citep{sutton1991sunk}, spatially varying markups would tend to produce an estimated tax that falls with market thickness. Section~\ref{se:corrRT} shows the opposite: the estimated tax is strongly positively correlated with population density. As an additional check, Appendix~\ref{ap:comp} conditions on the number of buildings constructed in the vicinity over the sample period. We find only a weak relationship, of inconsistent sign, between nearby construction and the estimated regulatory tax. Together, these findings suggest that spatially varying markups are unlikely to be a first-order driver of our results.

\subsubsection{Time and Space Varying Costs} \label{se:cost}

We also assume firms share the same non-land costs over space and time. This matters for identification because the frontier is interpreted as the unregulated cost at each height: if non-land costs differ across locations or time in ways not captured by our adjustments, then the inferred regulatory tax could reflect such cost differences rather than regulation.

To address non-land cost changes over time, we adjust prices using the Israeli Central Bureau of Statistics' residential construction input-prices index.\footnote{Estimates without adjusting for construction cost changes are similar (see Figure \ref{fig:mlerobust3}).}
Regarding spatial variation in non-land costs, industry participants suggest that such differences are small relative to the price differences we study.\footnote{Industry participants point out two variations: the cost of protecting the underground portion of very tall buildings from water encroachment in Tel Aviv and potentially lower labor costs in the Beer Sheva administrative tax region. These interviews were conducted for \cite{genesove2020}. \label{footnoteindustry}} This is corroborated by similar frontier estimates on samples that remove areas known to face greater technical challenges (see Figure \ref{fig:mlerobust2}).

\subsubsection{Differential Firm Efficiency} \label{se:firmeff}

Distinct from location and time, firms may also differ in their technical efficiency. If so, then the frontier costs are those of the most efficient firms operating  with the lowest markup in the least regulated market, and deviations from that frontier would reflect regulation as well as any remaining cost wedge or efficiency differences.\footnote{This is in the spirit of \cite{sutton1991sunk}, who in estimating the lower envelope of concentration ratios across normalized market sizes assumes a positive probability of maximally competitive conditions. Note also that the spatial component of the lower bound for the regulatory tax in Section \ref{se:bounds} can accommodate a minimum wedge that is weakly complementary with spatial amenities in the same manner as housing quality.}  

Yet the same diagnostics as for the importance of varying markups in Section \ref{se:pricetaking} are relevant here. Not only markups but firm differential inefficiencies as well should decrease with market size \citep{syverson2004market}.  Section~\ref{se:corrRT} shows that the estimated regulatory tax is highly positively correlated with population density, and Appendix~\ref{ap:comp} shows only a weak economic relationship, of inconsistent sign, between construction and the estimated regulatory tax. Together, these patterns suggest that firm differences are unlikely to be a major source of our regulatory tax estimates.

\subsubsection{Support} \label{se:support}

We have assumed a positive probability of observing unregulated markets at heights for which there are diseconomies of scale, and regulated markets at heights for which there are economies of scale, in place of the standard exogeneity assumptions for identification.
The frontier is not identified if these markets are not realized.  
Of course, there can be no hope of uncovering costs in the absence of regulation that is always imposed, such as  nationwide safety regulations.  Thus ``unregulated" should really be interpreted as ``minimally regulated", and it is the ``minimally regulated" frontier that is our estimation objective.
The problem arises rather when minimal regulation is realized at certain heights, but not at others.  However, that scenario might be detectable if one ends up estimating a nonsensical cost function. For example, consider transaction prices from a period of stable prices in a locality well characterized by the demand conditions of the monocentric city model, where willingness to pay decreases from the city center.  A greenbelt, where construction is forbidden, that surrounds the city would leave no way to identify marginal costs for heights that would have otherwise been built there. In this case, identification failure for this part of the frontier supply would be apparent from the gap in the distribution of prices, unconditional on height, whereas multiple time periods with changing overall prices will introduce more unregulated height away from the greenbelt, thus restoring identification. 

\subsubsection{No Subsidization and Price Expectations} \label{se:p=mc}

Below-cost prices would undermine frontier identification by violating the assumption that the regulatory tax is nonnegative.  Below-cost prices can be due either to government subsidization of construction costs, forced building beyond profit-maximizing heights or expectation mistakes.  Although there have been periods of government subsidization of construction costs, notably in response to the mass immigration from the ex-Soviet Union of the early 1990s \citep{genesoveFalk}, these were absent during our period of analysis.  

If builders expect a higher apartment price than what materializes, price may not cover cost.  We do not think this is a major concern, however.  Building-specific expectation mistakes can be included in measurement error:  under rational expectations, the observed price is a random deviation from the expected price, which is the relevant price for determining the cost frontier.  As modeled, however, measurement error fails to cover market-wide misperceptions. This should not be an issue, however, as parsimonious models forecast prices over the sample period fairly well.  A yearly AR(1) specification with a trend and structural break in trend at 2009 yields a root mean squared error of 0.018.\footnote{Housing prices rose steeply after the Bank of Israel drastically reduced interest rates at the beginning of 2009, as part of the coordinated, worldwide central bank response to the financial crisis.  Unanticipated price increases do not threaten identification of the frontier.}  Also, we do not see large variation in mean price differences across transactions within buildings that take place the year before, the year of or the year after construction, as we would expect to see if substantial surprises were common. Finally, when repeating our estimation on the pre-2008 period only, a period of relatively stable housing prices, we get similar results (see Figure \ref{fig:mlerobust2}).

	\section{Estimation}\label{se:est}

	\subsection{The Model}	\label{se:est_model}

	Consider the log prices of apartments in buildings of height $h$,
	\begin{align}
		y_{kij} 
		&= g  +u_{k} +w_{ki} + v_{kij} , && k=1,\ldots,K, \ i=1,\ldots,n_{k}, \ j=1,\ldots,J_{ki}, \label{ys}
	\end{align}
where $y_{kij}$ is the observed log price per square meter of apartment $j$ in building $i$ in bloc $k$, $g$ is the frontier,
$u_k$ is the deviation from the frontier,
	$w_{ki}$ is building-level measurement error,
	and $v_{kij}$ is apartment-level measurement error.\footnote{The log price is $y=\ln (P)=\ln (G+U)=\ln [G(1+U/G)] \approx \ln G + U/G \equiv g + u$.}
	 The distributions of $u_k\in[0,\infty)$, $w_{ki}\in(-\infty,\infty)$, and $v_{kij}\in(-\infty,\infty)$ can depend on height, but the lower bound of $u_k$ and the means of $w_{ki}$ and $v_{kij}$ are all equal to zero independently of height.\footnote{Spatial dependence of $u$ is considered in the robustness section.} 

	The conditional mean of \eqref{ys} is,
	\begin{align}
		E[y|h]&=g(h)+E[u|h], \label{Ey}  
	\end{align}
	as $E[w|h]=E[v|h]=0$ by assumption. Equation \eqref{Ey} demonstrates the importance of 
	having the parameters of the distribution of $u$ 
	depend on $h$. 
	Were these parameters, instead, the same across heights, then frontier estimates would equal the height-specific means, up to a common constant, making frontier analysis pointless.  Further, in this case, any endogeneity bias present in conditional mean analysis would also be present here.
		Hence, $u$ (and $v$ and $w$) are allowed to have separate parameters for each height.
		However, $u$'s distribution originates in the joint distribution of demand and supply shocks  through the equilibrium condition. Thus, unlike frontier costs $g(h)$, the parameters of $u$'s distribution will not be ``deep parameters." 
	
\subsection{Variances} \label{se:decompvar}
	
Without invoking any distributional assumptions, we identify and estimate the variances of $u$, $v$, and $w$ using the  hierarchical structure (formulas are in Appendix~\ref{ap:var}; see \citealp{BenMosheGenesoveMoments} for derivations).
Specifically, conditional on height $h$, the variance of the apartment-level measurement error $v$ is identified by within building variation in apartment time-adjusted prices,
 the variance of the building-level measurement error $w$ is identified by within bloc variation in building time-adjusted prices,
 and the variance of the deviations $u$  is identified by variation in prices (unadjusted for time) across both bloc and time.

\subsection{The Frontier}\label{se:frontest}

We estimate the frontier by maximum likelihood.\footnote{We have considered alternative estimators.  The commonly used, and convenient, priors of  Bayesian  estimators are not readily compatible with a frontier objective, while minimum-price-adjusted estimators converge slowly at logarithmic rates \citep[see][]{goldenshluger2004bndry}.}
At height $h$, assume that $v_{kij}\sim N(0,\sigma_v^2(h))$ and $w_{ki}\sim N(0,\sigma_w^2(h))$ are normal
and that $u_{k}\sim TN(\mu_u(h),\sigma_u^2(h))$ is the normal distribution truncated from below at zero.\footnote{If $x\sim N(\mu_x,\sigma_x^2)$ then $x \ | \ a \leq x<b$ is truncated normal. Although the truncated normal is not new to the SFA literature, the half-normal distribution (i.e., $\mu_x=0$) is more commonly used \citep[e.g.,][]{cai2021wrong}. However, this assumes deviations from the frontier are clustered near it, which we do not find in general.} 

Our approach contrasts with estimation based only on a cross-section of non-hierarchical data, where identification relies on the asymmetry of $u$ together with the symmetry of $v$ and $w$ to disentangle the deviation from the random errors. In our framework, identification of $\Var(u)$, $\Var(w)$, and $\Var(v)$ relies only on the hierarchical variance decomposition (see Section \ref{se:decompvar}). We then adopt truncated-normal/normal distributions for $u$, $w$, and $v$, following canonical SFA specifications, to obtain a tractable likelihood. We assess sensitivity to the assumed shape of $u$ with alternative distributions for $u$ (censored, folded, and two-sided symmetrically truncated normal), which yield similar frontier estimates (see Section \ref{se:robust}). In particular, estimates under symmetric and asymmetric specifications for $u$ are similar, mitigating concerns that observed right-skewness in prices is mechanically attributed to the regulatory tax.\footnote{Indeed, we find that the absolute value of the skewness of $u$ is below 0.5, which is considered small, at most heights.} 

The global maximum of the log likelihood, constrained so that average cost decreases to MES and marginal cost increases thereafter,  is attained by grid search and Dijkstra's algorithm,
\begin{align}
	&	\{\widehat{MES},\widehat g,\widehat \mu_u\} =\argmax_{
		\substack{mes \in \{1,\ldots,H-1\}   \\
							\textsl{g}\in \mathbb{R}^H,
							\nu_u \in \mathbb{R}^H}}  
						\	\sum_{h=1}^H\mathcal{L}_h(\textsl{g}_h,\nu_{uh},\cdot )  ,\label{eq:MLE} \\
		&\text{s.t. }  \textsl{g}_{mes} \leq \textsl{g}_{mes-1} \leq \ldots \leq \textsl{g}_1
		\textrm{ and }	\textsl{g}_{mes} \leq \textsl{g}_{mes+1} \leq \ldots \leq \textsl{g}_H,  \label{eq:MLE1} 
\end{align}
where $\mathcal{L}_h(\textsl{g}_h,\nu_{uh},\cdot )$ is the log likelihood at height $h$ (see Appendix \ref{ap:frontest} for details and formulas). The constraint allows for $\widehat{MES}=1$ and so no economies of scale.\footnote{We also present estimates that maximize the log likelihood at each height without constraints.}

		\subsection{Regulatory Tax Rates} \label{se:estRT}
	
  This section describes how to estimate and bound expected regulatory tax rates of error-free prices.
	Using the distributions from Section \ref{se:frontest} that $u\sim TN(\mu_u,\sigma_u^2)$ and $\eta \sim N(0,\sigma_\eta^2)$,
	 where $\sigma_\eta^2=\sigma_w^2+\sigma_v^2/J$ for building price and $\sigma_\eta^2=\sigma_w^2+\sigma_v^2$ for apartment price, we get,\footnote{Appendix \ref{ap:est} derives the conditional density when $u$ is truncated normal. \cite{jondrow1982estimation} derive the conditional density for the half-normal, which is the truncated normal with $\mu_u=0$.}
{\small\begin{align}
	u| u+\eta=y-g &\sim TN\Big(\frac{\mu_u\sigma_\eta^2+( y-g)\sigma_u^2}{\sigma_u^2+\sigma_\eta^2},
	\frac{\sigma_u^2\sigma_\eta^2}{\sigma_u^2+\sigma_\eta^2} \Big). \label{eq:distu1}  
\end{align}
}
Assuming that deviations from the frontier are entirely due to regulatory restrictions (taking into account the discreteness of height), the expected regulatory tax rate based on \eqref{RT} is, 
\begin{align}  
 	E[\frac{1}{G(h)e^{u}}\text{RT}(G(h)e^{u},h) |  y-g(h)]. \label{eq:RTRu}
\end{align}
where $u$ is drawn from \eqref{eq:distu1}, conditioned on $y_i-g(h_i)$.
However, if quality differs systematically over location then deviations also include quality. 
In this case, the lower bound based on \eqref{eq:RTbound} is,
\begin{align}
 & E\Big[\frac{1}{G(h_i)e^{u_i}} \cdot \min_{\kappa_{Si}\in[0,1]}\max_{j \in \Omega_i(d)}  \max\{0, G(h_j) - G(h_i+1)- (G(h_j)e^{u_j}-G(h_i)e^{u_i}) 	+\kappa_T (1-T_{ij})G(h_j)e^{u_j} \nonumber\\
& \hspace{3.0cm}
		+ \kappa_{Si} (T_{ij}G(h_j)e^{u_j}-G(h_i)e^{u_i}) \}
		\Big| y_i-g(h_i), y_j-g(h_j), j \in \Omega_i(d)\Big] , \label{eq:RTboundest} 
	\end{align}
	where $u_i$ and $u_j$, for $j\in \Omega_i(d)$, are drawn independently from \eqref{eq:distu1}, conditioned on $y_i-g(h_i)$ and $y_j-g(h_j)$.

\section{Data} \label{se:data}

Apartment transaction data are obtained from CARMEN, the digitalized repository of buyer reports to the Tax Revenue Authority.
The data include the transaction date, price, square meters, apartment floor, number of floors in the building, and year of construction.  They also include a unique identifying number from the land registry for the bloc and parcel on which the building sits, where the parcel is a lower level geographical division than the bloc, one or more of which comprise a single bloc.  In general, the building and bloc-parcel are coincident.  However, 534 buildings, or 2.9\% of buildings in the pre-exclusion sample, are located on 261 parcels containing more than one building. We exploit these cases to identify the hedonic height effects presented in Section \ref{se:floorheight} and estimated below in Section \ref{se:preresults}, but exclude them from the stochastic frontier analysis.
The sample covers the period 1998 to 2017.

We limit the sample to transactions from CARMEN for which (1) the year of
the transaction is the year before, the year of or the year after the construction
year, (2) the transaction is for 100\% of the asset, (3) the property type is not a single family home,  
(4) none of the variables listed above is missing, and (5) there is at least one other transaction observed in the building.
We adjust prices for apartment floor-space area by expressing them in per square meters. 
To account for inflation, we convert prices to real 2017 values.  These prices are adjusted for floor and height premia, as described in Section \ref{se:floorheight}. To estimate the frontier and regulatory tax, we further adjust for changes in construction input prices (other than land) over time by dividing the real prices by the Israeli Central Bureau of Statistics' residential construction input prices index, expressed in 2017 values. In addition, we exclude apartments with missing prices and apartments on parcels containing more than one building, removing about 4.4\% of the original observations.\footnote{We drop apartments with nominal prices in the bottom one percent and top one percent of the distribution. We also examined buildings that were low outliers conditional on height, with the lowest building price at least 10 percent below the next-lowest building price at that recorded height. This led us to exclude four buildings whose recorded heights were contradicted by multiple independent records. A small number of additional buildings have valid price and height information but coordinates inconsistent with the rest of their locality or bloc; we retain these observations but set their coordinates to missing for geographic analyses.}

 There are 2,449 blocs, 18,171 buildings, and 270,684 apartments in the resulting sample.\footnote{Table \ref{tablesumobs} in Appendix \ref{ap:tableobs} shows summary statistics for the number of observations by height.}
The median bloc size is about 0.21km$^2$.
Unconditional on height, the mean number of buildings in a bloc is about 7.5 in our transactions data. A bloc containing buildings of multiple heights therefore appears in multiple rows.
 
 Table \ref{tab:summary} shows apartment-level summary statistics of price (per square meter in real 2017 NIS and adjusted for cost) and the number of floors in the building (i.e., height),
and building-level summary statistics of price (average price within a building) and the number of floors in the building.
The mean real, input-price, height and floor-adjusted per square meter price is such that a standard 100 square meter apartment would sell for about 1.25 million NIS in 2017 shekels (about 350,000 USD at 2017 exchange rates).

\begin{figure}[tp]
	\centering
	\includegraphics[width=\linewidth]{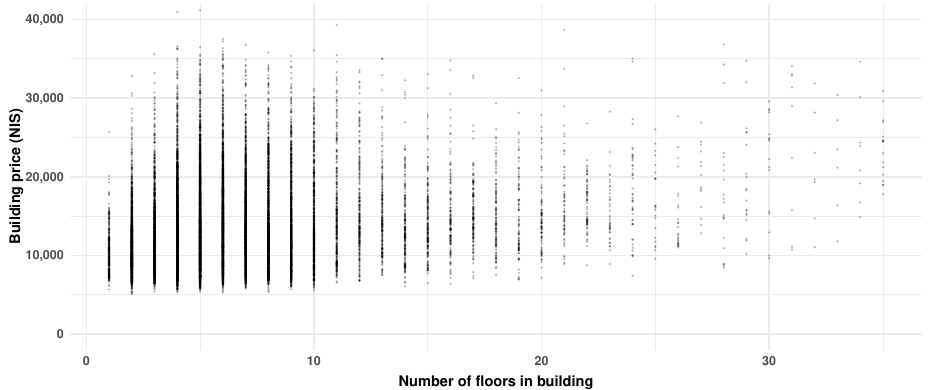}
	\caption{\footnotesize \hspace{-0.3cm}  Frequency of building prices in NIS (rounded to nearest 100) by height. \label{fig:prices}}
\end{figure}

The points in Figure \ref{fig:prices} are building prices  by height.
There is a large dispersion in prices at nearly all heights, with the average ratio of third to first quartile price equal to 1.6 and
the 95\% to 5\% price ratio equal to 2.7.

\captionsetup{width=\textwidth,justification=raggedright}

 \begin{table}[!htb]
	\centering
	\begin{threeparttable} 
		\caption{Summary statistics}
		\label{tab:summary} 
\begin{tabular}{@{}lrrrrr@{}}
\toprule
& Mean & Std. Dev. & Min & Med & Max \\
\midrule
\multicolumn{6}{@{}l}{\textit{Apartment}} \\
Log price        & 9.35   & 0.38   & 8.40   & 9.34   & 10.53   \\
Price            & 12,369 & 5,056  & 4,457  & 11,423 & 37,371  \\
Number of floors & 9.36   & 5.87   & 1      & 8      & 40      \\
\addlinespace
\multicolumn{6}{@{}l}{\textit{Building}} \\
Log price        & 9.36   & 0.39   & 8.49   & 9.35   & 10.50   \\
Price            & 12,529 & 5,205  & 4,852  & 11,461 & 36,329  \\
Number of floors & 6.65   & 4.51   & 1      & 6      & 40      \\
\bottomrule
\end{tabular}

	\end{threeparttable}
	
\vspace{4pt}
	\begin{minipage}{\textwidth}
		\footnotesize
		\textit{Notes:} Prices per square meter in real 2017 NIS. There are 18,171 buildings and 270,684 apartments.
	\end{minipage}
\end{table}

\FloatBarrier
\section{Results} \label{se:results}

\subsection{Apartment-Floor, Building-Height Adjusted Price} \label{se:preresults}

Adjusting prices for observable attributes is especially important in our context.  
On the one hand, consumers may be prepared to pay a premium, or demand a discount, for apartments on high floors or in tall buildings.  On the other hand, building height varies with location, with taller buildings constructed in more attractive areas, as basic land use theory predicts.  
The challenge is to obtain an empirical counterpart to $p$ of \eqref{Price}, the price after removing apartment-floor, building-height effects. An insufficiently flexible specification could easily assign  apartment floor or building height effects to location effects, thus overstating the increase in the frontier at higher heights; too much flexibility could lead to excessive noise in the estimates.  Our solution is to first estimate a fully saturated model of floor and height effects, and then, after inspecting the estimates, choose a reasonable restricted model.
The function $m$ in \eqref{Price} is identified using variation in apartment floor within a building and variation in building height within a parcel, as some parcels have more than one building on them.\footnote{See Appendix \ref{ap:PSR} for details. We normalize $m(2,4) = 1$, so that the adjusted price represents a second-floor apartment in a 4-floor building at the given location.} 
We then subtract the estimated floor and height effects from the observed price and add back in the effects pertaining to a second-floor apartment in a 4-floor building.  This is the price used in the remainder of the analysis.

\subsection{Variances} \label{se:varresults}

\begin{figure}[tp]
	\centering
	\includegraphics[width=\linewidth]{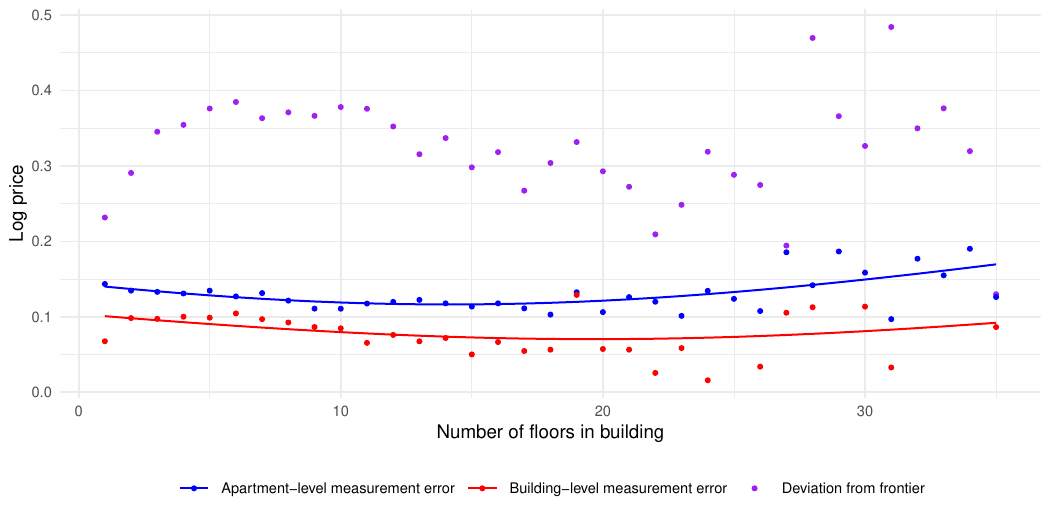}
	\caption{\footnotesize \hspace{-0.3cm} The red, blue, and purple points are estimated standard deviations based on \eqref{varv2}–\eqref{varu2}. The red and blue curves smooth the estimates with series estimators.}
	\label{fig:sd}
\end{figure}

Figure \ref{fig:sd} shows the estimated standard deviations, by height, of apartment-level measurement error $v$ (in blue), building-level measurement error $w$ (in red), and deviations from the frontier $u$ (in purple),
using \eqref{varv2}-\eqref{varu2} in Appendix \ref{ap:var}. The measurement error variances are estimated using residuals of a nonparametric
	regression of log price on transaction day. The deviations variance is then estimated using log prices and the estimated measurement error variances. Thus the variance of deviations ($\approx$ regulations) is obtained from variation in prices (unadjusted for time) across both bloc and time, while the variances of measurement errors partial out time effects.
For some of the higher heights, the degrees of freedom at the building-level are small or zero (see Table \ref{tablesumobs} in Appendix \ref{ap:tableobs}) so that the estimated building-level measurement error variances do not exist or are negative, and so are missing from the figure.  To deal with these cases and to avoid excessively noisy estimates, we smooth the measurement error variances
using polynomial series estimates, with the polynomial degrees chosen by cross validation. The resulting  curves are relatively flat. 
We do not smooth the standard deviations of $u$. 
Allowing these standard deviations to be unrestricted functions of height avoids imposing any endogeneity bias, as we discussed underneath \eqref{Ey}.

\captionsetup{width=\textwidth,justification=raggedright}

The figure shows that the estimated standard deviation of $u$ is on average about 4 times the estimated standard deviation of building error and about 2.5 times the estimated standard deviation of apartment error.
Thus the variance of regulation is several times larger than the combined measurement error variance. The standard deviations of the measurement errors, however, are clearly nontrivial.

\FloatBarrier
\subsection{The Frontier} \label{se:frontresults}

\begin{figure}[tp]
	\centering
	\includegraphics[width=\linewidth]{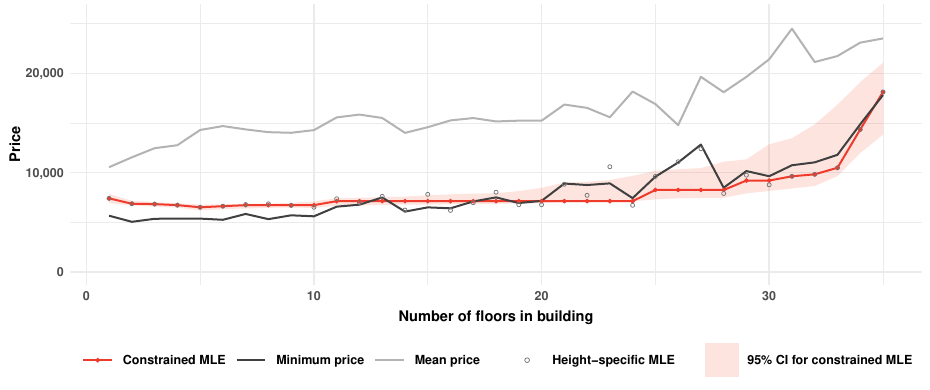}  
	\caption{\footnotesize \hspace{-0.3cm} The minimum and mean building prices, constrained MLE, and height-by-height MLEs. The shaded region is the 95\% confidence band from 200 parametric bootstrapped samples; hollow circles are MLEs estimated separately at each height.}
	\label{fig:mleCI}
\end{figure}

Figure \ref{fig:mleCI} shows our constrained frontier maximum likelihood estimate (MLE) from \eqref{eq:MLE}-\eqref{eq:MLE1} (see also Appendix \ref{ap:mlenum}), together with estimates obtained separately at each height. The estimates decrease until MES at five floors, remain relatively constant, and then increase steeply.  
Although the upper confidence band permits marginal costs that rise immediately beyond MES, each parametric bootstrap sample produced a frontier with long stretches of constant marginal cost.\footnote{Let $(\widehat g(h), \widehat \sigma_v^2(h),\widehat \sigma_w^2(h),\widehat \sigma_u^2(h),\widehat \mu_u(h))$ be the MLE. The parametric bootstrap at height $h$ randomly draws  $v_{kij}^*$ from $N(0,\widehat \sigma_v^2(h))$, $w_{ki}^*$ from $N(0,\widehat \sigma_w^2(h))$, and $u_{k}^*$ from $TN(\widehat \mu_u(h),\widehat \sigma_u^2(h))$. The bootstrapped observation is $y^*_{kij}=\widehat g(h)+ u_{k}^* +w_{ki}^* +v_{kij}^*$.} The figure also shows mean and minimum building prices. The differences between mean prices and the MLE, along with the relative sizes of the variances estimated in Section \ref{se:varresults}, show that multi-floor housing markets must be highly regulated, with some building prices more than six times frontier prices. A difference between mean prices and the MLE is that the former increase sharply at low heights but the latter  decrease. Minimum prices are consistent estimators for the frontier absent measurement error (see Section \ref{se:supply}) but with measurement error, at low heights, where there are many buildings with data on just two apartments, it is likely that some building has large negative measurement error and is relatively unregulated, making minimum prices biased downwards as frontier estimates. At high heights, there are relatively few buildings and so minimum prices will tend to be biased upwards as frontier estimates.  The average cost at MES is about 12\% lower than the average cost of constructing a one-floor building. The marginal cost initially increases, then remains flat, before increasing steeply, reflecting that building upwards becomes increasingly difficult at high heights. This is consistent with previous research \citep[e.g.,][]{glaeser2005manhattan} and discussions with industry experts (see footnote \ref{footnoteindustry}).

\captionsetup{width=\textwidth,justification=raggedright}

Table \ref{table:regfront} compares buildings near the frontier, defined as buildings with average apartment price at most 5\% greater than the frontier, to the full sample of newly constructed buildings.\footnote{Table \ref{table:regfront} and the analysis in Section \ref{se:corrRT} use the subset of the data with geographical coordinates.} 
About 3.5\% of apartments in the full sample are near the frontier. Relative to the full sample, housing near the frontier is about twice as far from the city of Tel Aviv, the country’s commercial center.  Depending on the radius and whether we look at buildings or apartments, `Near Frontier' housing is in areas with average densities, in 10,000's per km$^2$, between 0.31 and 0.62 that of the full sample.  The smaller standard deviations for `Near Frontier' indicate greater homogeneity of this sub-sample relative to the full sample.  Although these buildings are further away from Tel Aviv, they are, perhaps surprisingly, closer to their own locality centers,  but the standard deviation indicates a large degree of disparity. 

\begin{table}[!htb]
	\centering
	\begin{threeparttable} 
		\caption{Comparison of full sample and near frontier}
		\label{table:regfront} 
\begin{tabular}{@{}lrrrr@{}}
\toprule
& \multicolumn{2}{c}{Full sample} & \multicolumn{2}{c}{Near frontier} \\
& Mean & Std. Dev. & Mean & Std. Dev. \\
\midrule
\multicolumn{5}{@{}l}{\textit{Apartment}} \\
Regulatory tax rate                    & 0.45 & 0.15 & 0.12 & 0.04 \\
Distance to locality center            & 2.43 & 1.56 & 1.89 & 1.22 \\
Density (1km radius)                   & 0.50 & 0.50 & 0.31 & 0.27 \\
Density (4km radius)                   & 0.32 & 0.27 & 0.14 & 0.14 \\
Distance to Tel Aviv locality (km)     & 37.74 & 35.58 & 70.59 & 29.46 \\
\addlinespace
\multicolumn{5}{@{}l}{\textit{Building}} \\
Regulatory tax rate                    & 0.48 & 0.16 & 0.11 & 0.05 \\
Distance to locality center            & 2.43 & 1.57 & 1.82 & 1.32 \\
Density (1km radius)                   & 0.62 & 0.57 & 0.26 & 0.22 \\
Density (4km radius)                   & 0.41 & 0.34 & 0.13 & 0.12 \\
Distance to Tel Aviv locality (km)     & 37.89 & 38.50 & 79.96 & 28.70 \\
\bottomrule
\end{tabular}

	\end{threeparttable}
	
	\vspace{4pt}
	\begin{minipage}{\textwidth}
		\footnotesize
		\textit{Notes:} We remove observations with missing geographical coordinates so that there are
		13,102 buildings and 206,822 apartments in the full sample and 
		350 buildings and 7,215 apartments near the frontier.
		Distances are in kilometers. Densities are in units of 10,000 per km$^2$.
	\end{minipage}
\end{table}

Consistent with our general view of regulatory variation as extremely local, buildings near the frontier are well represented throughout the country, with 59 of the 160 localities in Table \ref{table:regfront} having at least one building near the frontier. 
Seven administrative tax regions contain over 99\% of buildings near the frontier. The remaining three regions are those closest to Tel Aviv.

\FloatBarrier
\subsection{Robustness of the Frontier} \label{se:robust}

In our primary analysis, $u_k$, representing deviations from the frontier, follows a truncated normal distribution. 
To test the sensitivity of our results to this assumption, we considered alternative distributions for $u_k$.  

We first consider a folded normal -- like our baseline truncated normal, this is a generalization of the half-normal, the original and still often used specification for the deviation in the SFA literature.  We also use a zero-censored normal, although this assumes a prevalence of minimally regulated buildings, which is inconsistent with our baseline estimates, where the mean of $u_k$ is often much larger than its variance. Finally, we also estimate a two-sided, symmetrically truncated normal (with support $[0,2\mu]$); here identification relies only on the hierarchical structure (pre-estimated variance components for $u$, $v$, and $w$) and not on skewness in any way. We also report a method-of-moments truncated-normal estimate using the pre-estimated variance components. As shown in Figure~\ref{fig:mlerobust1}, our estimates are robust to these different distributional and estimation choices. Because low-price observations identify the frontier in the absence of measurement error, we also verified that the qualitative frontier shape is unchanged when dropping the lowest-priced building at each height. 

To assess robustness to spatial dependence, we specify the spatial autoregressive relationship $u_k = \rho \sum_{l=1}^K \omega_{kl} u_l + \zeta_k$. As shown in Figure~\ref{fig:mlerobust2}, the overall shape of the frontier is similar under spatially correlated deviations, though the spatial estimate lies below the baseline.

We also considered building-level regulations, modifying the model to $y_{kij}=g+u_{ki}+w_{ki}+v_{kij}$. Identification now depends on the skewness of the distribution of $u_{ki}$ and the symmetry of the distribution of $w_{ki}$. The practical application of this model requires $\sigma_u^2$ to be sufficiently larger than $\mu_u$ for the distinction between a truncated normal distribution and a normal distribution to be discernible. Figure~\ref{fig:mlerobust3} shows similar estimates using building-level regulations.

To assess robustness to spatial cost differences, we re-estimated the frontier excluding the Beer Sheva tax assessment area, where labor costs may be lower. Figure~\ref{fig:mlerobust2} shows that the estimated frontier is robust across different spatial contexts. Excluding Tel Aviv, where building near the aquifer raises costs at large heights, has no effect because there are no observations near the frontier. We also examined robustness to temporal cost differences by estimating the frontier without time adjustments and, separately, by restricting the sample to the pre-2008 period of relatively stable housing price growth. Figures~\ref{fig:mlerobust1} and \ref{fig:mlerobust2} show that the estimates are stable over time.

\begingroup
\setlength{\textfloatsep}{5pt plus 1pt minus 1pt}
\setlength{\floatsep}{5pt plus 1pt minus 1pt}
\captionsetup{width=.94\textwidth,justification=raggedright,font=footnotesize}
\captionsetup[subfigure]{font=footnotesize,justification=raggedright,
	singlelinecheck=false,skip=2pt}

\begin{figure}[tp]
	\centering
	\begin{adjustwidth}{-0.8em}{-0.8em}
		\makebox[\textwidth][c]{%
			\begin{subfigure}[t]{0.335\textwidth}
				\centering
				\includegraphics[width=\linewidth,height=0.22\textheight]{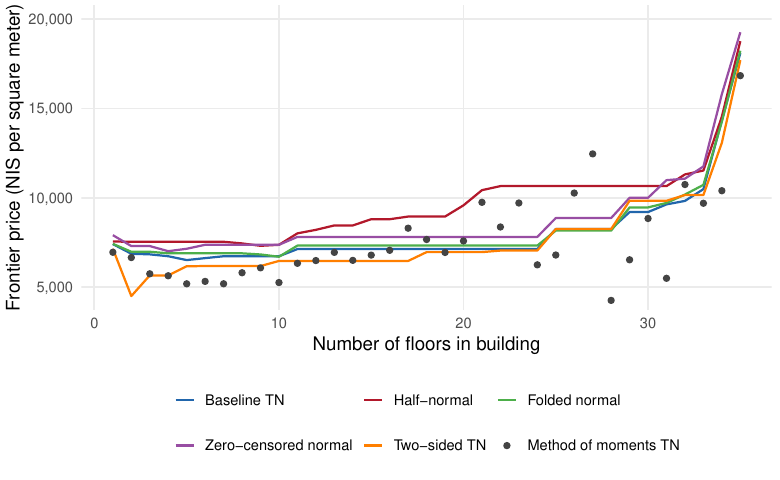}
				\vspace{-0.4em}
				\subcaption{\footnotesize Robustness to censored normal, folded normal, two-sided truncated normal, and MM with truncated normal (by height).}
				\label{fig:mlerobust1}
			\end{subfigure}\hspace{0.012\textwidth}%
			\begin{subfigure}[t]{0.335\textwidth}
				\centering
				\includegraphics[width=\linewidth,height=0.22\textheight]{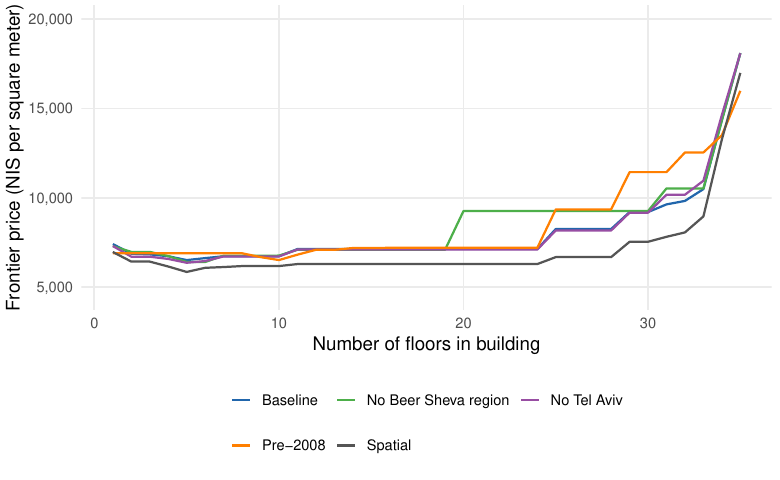}
				\vspace{-0.4em}
				\subcaption{\footnotesize Robustness to excluding the Beer Sheva administrative tax region, using pre-2008 data, and spatial correlation.}
				\label{fig:mlerobust2}
			\end{subfigure}\hspace{0.012\textwidth}%
			\begin{subfigure}[t]{0.335\textwidth}
				\centering
				\includegraphics[width=\linewidth,height=0.22\textheight]{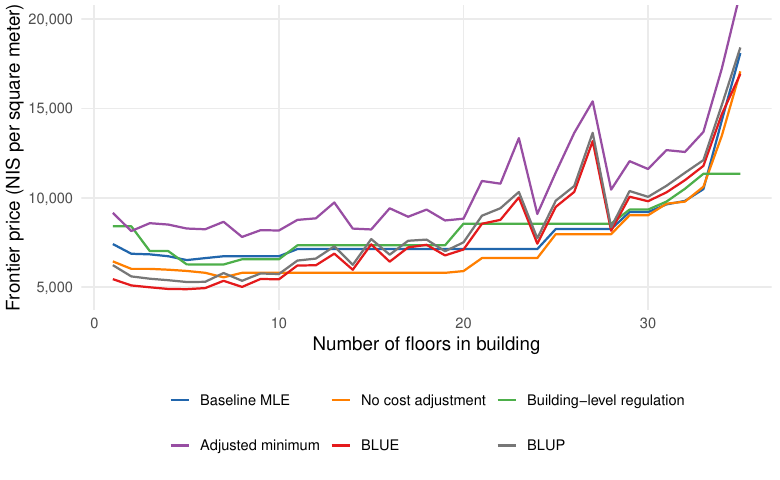}
				\vspace{-0.4em}
				\subcaption{\footnotesize Robustness to not adjusting for changes in cost over time, building-level regulation, adjusted minimum price, and BLUE.}
				\label{fig:mlerobust3}
			\end{subfigure}%
		}
	\end{adjustwidth}
	
	\caption{\footnotesize Frontier robustness: spatial/temporal variants, estimators, and distributional assumptions.}
\end{figure}
\endgroup

Additionally, we employed the best linear unbiased estimator (BLUE) and the best linear unbiased predictor (BLUP), with uninformative and normal priors on $g+u_k$, respectively. While these approaches are conventionally used for mean estimation, in our case they are less suitable since $g+u_k$ represents a minimum. These estimates of the frontier, shown in Figure~\ref{fig:mlerobust3}, are similar to our frontier MLE.

Lastly, we estimated the frontier by a sample-size adjustment to the minimum price, as proposed by \cite{goldenshluger2004bndry}. This involved estimating the frontier as $\widehat g_{\text{GTm}}=\min_{k,i}\{\tfrac{1}{m}\sum_{j=1}^m y_{kij}\}+\widehat \sigma_{GTm}\sqrt{2\ln(n)}$. The results, illustrated in Figure~\ref{fig:mlerobust3}, have a similar shape to our estimates but are substantially higher, perhaps due to slow log convergence rates.

These robustness checks, covering distributional assumptions, spatial and temporal variations, and alternative estimation techniques, support the validity of our frontier estimates.

\FloatBarrier
\subsection{Substitution of Land for Capital} \label{se:EOS}

\begin{figure}[tp]
	\centering
	\includegraphics[width=0.75\linewidth]{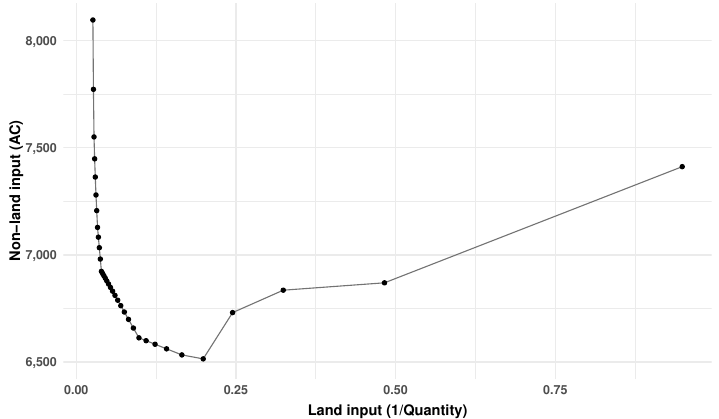}
	\caption{\footnotesize Frontier isoquant: non-land input (average cost) against land input (1/quantity).}
	\label{fig:eosiso}
\end{figure}

The cost frontier also describes how non-land inputs and land substitute in housing production. Figure \ref{fig:eosiso} plots the implied frontier isoquant, with non-land input per unit of housing (average cost) on the vertical axis and land input per unit (one over quantity) on the horizontal axis. The increasing segment at high land use (low heights, up to MES) reflects economies of scale. Over roughly 11 to 24 floors, marginal cost is nearly constant, corresponding to an approximately linear segment of the isoquant. At high heights the isoquant becomes steep, so that additional floors require substantially more non-land input for little saving in land. Substitution is thus easiest over the middle range, where the structural systems needed to build taller are already in place, and most limited at the highest heights.

Because building height is discrete in our setting, we quantify substitution using arc elasticities between land and non-land inputs. The elasticity of substitution is not defined below MES, is infinite where marginal cost is constant, and is sensitive at kinks where marginal cost jumps; at the highest heights, from 30 to 35 floors, the aggregate arc elasticity is about 0.15 to 0.2 depending on the differencing convention. Most of the literature estimates or assumes a constant elasticity, often near unity \citep[e.g.,][]{ahlfeldt2014new,combes2021production}, with the few estimates for tall residential buildings about 0.5 \citep{ahlfeldt2018tall}; our estimates suggest instead that substitutability varies substantially with height and is well below these values at the highest heights.

\subsection{Regulatory Tax Rates} \label{se:RTres}

In this section, we demonstrate that the regulatory tax is substantial by estimating its mean, standard deviation, and lower bounds. For each building we estimate the upper bound (based on \eqref{eq:RTRu}) and lower bounds (based on \eqref{eq:RTboundest}) for the mean regulatory tax rate. Recall that the upper bound is simply the point estimate.  
The lower bounds use nearby buildings within distance $d\in \{\text{1km, 2km, 3km, 4km}\}$.  
To recall, we restrict the set of comparison buildings used in the construction of the lower bound to those lying within a certain radius, both because our bound is based on a first order approximation and because we prefer to be conservative in restricting the distribution of preferences, that is, we do not assume that the distribution of preferences over quality and locational amenities is identical across all locations.
The mean number of buildings within 1km, 2km, 3km, and 4km is 80, 195, 315,  and 435 respectively.\footnote{Four kilometers is large enough to cover the entire locality in which a building is located in most cases.} The existing home price regression yields an estimate of 0.0016 for $\kappa_T$, as reported in Appendix \ref{ap:timeeffects}.  The estimated mean value of $\kappa_{Si}$ is $0.65$, with standard deviation $0.35$.  

Across all buildings, the mean and standard deviation of the upper bound are 47.4\% and 16.5\%.  
Thus the regulatory tax is not only high on average but also highly variable.
Across all buildings with heights above MES, the mean and standard deviation of the upper bound are 49\% and 17\%.\footnote{Across all apartments, the upper bound is 45\%, with a standard deviation of 15\%. Across all apartments in buildings with heights at or above MES (i.e., five floors), the upper bound is 46\%, with a standard deviation of 15\%.}
Restricting to buildings with geographical coordinates, and using buildings within 1km, 2km, 3km, and 4km respectively, the lower bounds are 19\%, 24\%, 28\%, and 31\% with standard deviations 16\%, 20\%, 23\%, and 24\%.
	Restricting further to buildings with heights above MES, the lower bounds are 23\%, 29\%, 33\%, and 37\% with standard deviations 17\%, 20\%, 23\%, and 25\%.
	
\begin{figure}[tp]
	\centering
	\includegraphics[width=\linewidth]{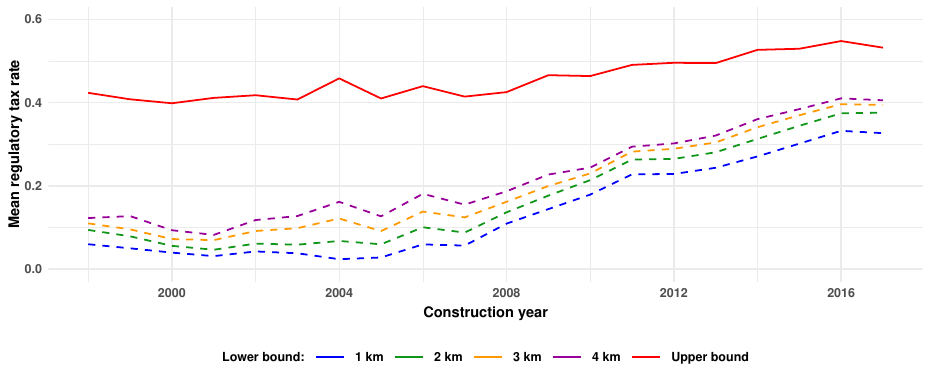}  
	\caption{\footnotesize \hspace{-0.3cm}  The upper and lower bounds for mean regulatory tax rates for buildings with heights from 6 to 30.}
	\label{fig:lbovertime}
\end{figure}

Figure \ref{fig:lbovertime} shows the upper and lower bounds over time for buildings with heights from 6 to 30 floors. 
The lower bounds tighten substantially with time as housing prices increased, post-2008. In 2017, with housing prices having approximately doubled since 2008, the lower bounds for the four radii reach 32.7\%, 37.6\%, 39.5\%, and 40.6\%, with the upper bound at 53.2\%.
With a small estimate for $\kappa_T$, this demonstrates the greater usefulness of bounds in periods that follow high price growth.

\captionsetup{width=\textwidth,justification=raggedright}

\FloatBarrier
\subsection{Characterizing Regulatory Tax Rates}\label{se:corrRT} 

We document the pattern of the estimated regulatory tax rate within and across localities. Our aim is solely descriptive. The results point to both broad locality-level differences and substantial within-locality heterogeneity.

\begin{figure}[tp]
	\centering
	\includegraphics[width=\linewidth]{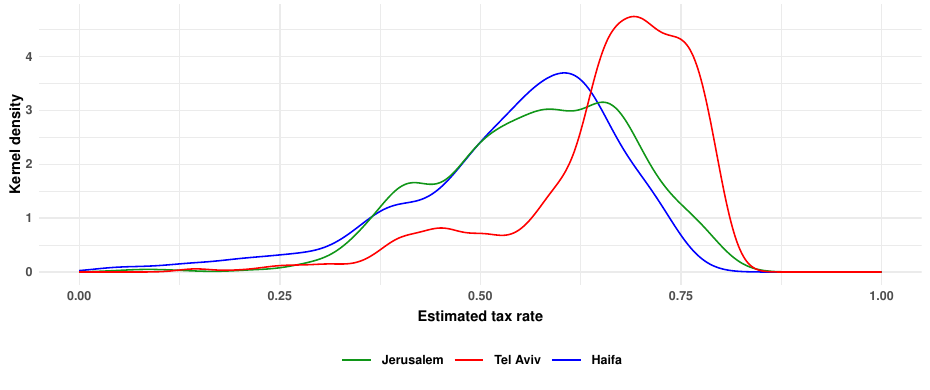}
	\caption{\footnotesize \hspace{-0.3cm}  The kernel densities of the estimated regulatory tax rates in Jerusalem, Tel Aviv, and Haifa.}
	\label{fig:citydist}
\end{figure}

Figure \ref{fig:citydist} shows the distributions of estimated regulatory tax rates for Jerusalem, Tel Aviv, and Haifa, showing both within-locality heterogeneity and across-locality differences. Regression (3) in Table \ref{table:regregs}, which includes only locality fixed effects, achieves an $R^2$ of 0.667 at the building level. This implies that a building’s locality is the strongest single predictor of its regulatory tax rate, explaining about two-thirds of the nationwide variation. Nevertheless, approximately one-third remains unexplained, implying within-locality heterogeneity in regulation.

We now examine the within-locality variation in the regulatory tax rate by correlating it with distance to the locality center and density. 
We define the locality center as the location within the locality with the highest predicted price from a nonparametric regression of building prices on geographical coordinates, with bandwidth chosen by cross-validation.
This definition is consistent with monocentric locality models and avoids relying on employment or amenity data.
Population density is measured  as the 1995 resident population within a 1km or 4km radius of each building.
Regressions (4), (5), (6), and (7) in Table \ref{table:regregs}, all estimated with locality fixed effects, show a negative correlation between distance to the locality center and regulatory tax rate, and a positive correlation between density and the regulatory tax rate. The negative correlation with distance to locality center is consistent with \cite{tan2020land}, who proxy the locality center using nighttime light intensity. The positive correlation with density is consistent with \cite{hilber2013origins}, who find that areas with higher land development tend to have stricter regulation, reflecting incumbent homeowners' preferences. While both relationships are statistically significant, they add little explanatory power beyond locality fixed effects: including them increases the $R^2$ in the building-level regression from 0.667 to 0.672.

\captionsetup{width=\textwidth,justification=raggedright}

\begin{table}[!htb]
	\centering
	\begin{threeparttable}
		\caption{Regressions}
		\label{table:regregs}
		\small
\begin{tabular}{@{}lccccccc@{}}
\toprule
\multicolumn{8}{@{}l}{\textit{Apartment}} \\
\midrule
\T Distance to locality center & \ \ - & \ \ - & \ \ - & \makecell {\ \ -0.004 \\ \footnotesize{(0.001)}} & \ \ - & \ \ - & \makecell {\ \ -0.005 \\ \footnotesize{(0.002)}} \\
\T Density (1km radius) & \makecell {\ \ 0.091 \\ \footnotesize{(0.004)}} & \ \ - & \ \ - & \ \ - & \makecell {\ \ 0.012 \\ \footnotesize{(0.004)}} & \ \ - & \ \ - \\
\T Density (4km radius) & \ \ - & \makecell {\ \ 0.278 \\ \footnotesize{(0.007)}} & \ \ - & \ \ - & \ \ - & \makecell {\ \ 0.063 \\ \footnotesize{(0.017)}} & \makecell {\ \ 0.070 \\ \footnotesize{(0.017)}} \\
\T Locality fixed effects & No & No & Yes & Yes & Yes & Yes & Yes \\
\T $R^2$ & \ \ 0.086 & \ \ 0.229 & \ \ 0.562 & \ \ 0.563 & \ \ 0.563 & \ \ 0.564 & \ \ 0.565 \\
\addlinespace
\multicolumn{8}{@{}l}{\textit{Building}} \\
\midrule
\T Distance to locality center & \ \ - & \ \ - & \ \ - & \makecell {\ \ -0.007 \\ \footnotesize{(0.001)}} & \ \ - & \ \ - & \makecell {\ \ -0.008 \\ \footnotesize{(0.001)}} \\
\T Density (1km radius) & \makecell {\ \ 0.104 \\ \footnotesize{(0.002)}} & \ \ - & \ \ - & \ \ - & \makecell {\ \ 0.016 \\ \footnotesize{(0.002)}} & \ \ - & \ \ - \\
\T Density (4km radius) & \ \ - & \makecell {\ \ 0.316 \\ \footnotesize{(0.004)}} & \ \ - & \ \ - & \ \ - & \makecell {\ \ 0.083 \\ \footnotesize{(0.010)}} & \makecell {\ \ 0.099 \\ \footnotesize{(0.010)}} \\
\T Locality fixed effects & No & No & Yes & Yes & Yes & Yes & Yes \\
\T $R^2$ & \ \ 0.126 & \ \ 0.305 & \ \ 0.668 & \ \ 0.670 & \ \ 0.669 & \ \ 0.670 & \ \ 0.673 \\
\bottomrule
\end{tabular}

	\end{threeparttable}
	
		\vspace{4pt}
	\begin{minipage}{\textwidth}
		\footnotesize
		\textit{Notes:} Standard errors are in parentheses underneath the coefficients. 
		Distance to locality center is in kilometers.
		Densities are in 10,000's per square kilometer. 
		There are 13,102 buildings and 206,822 apartments. 
		The top panel has outcomes at the apartment level, and standard errors clustered at the building level.
		The bottom panel uses the building-level mean apartment price.
	\end{minipage}
\end{table}

\begin{figure}[tp]
	\centering
	\begin{adjustwidth}{-1.2em}{-1.2em}
		\makebox[\linewidth][c]{%
			\begin{subfigure}{0.495\linewidth}
				\centering
				\includegraphics[width=\linewidth]{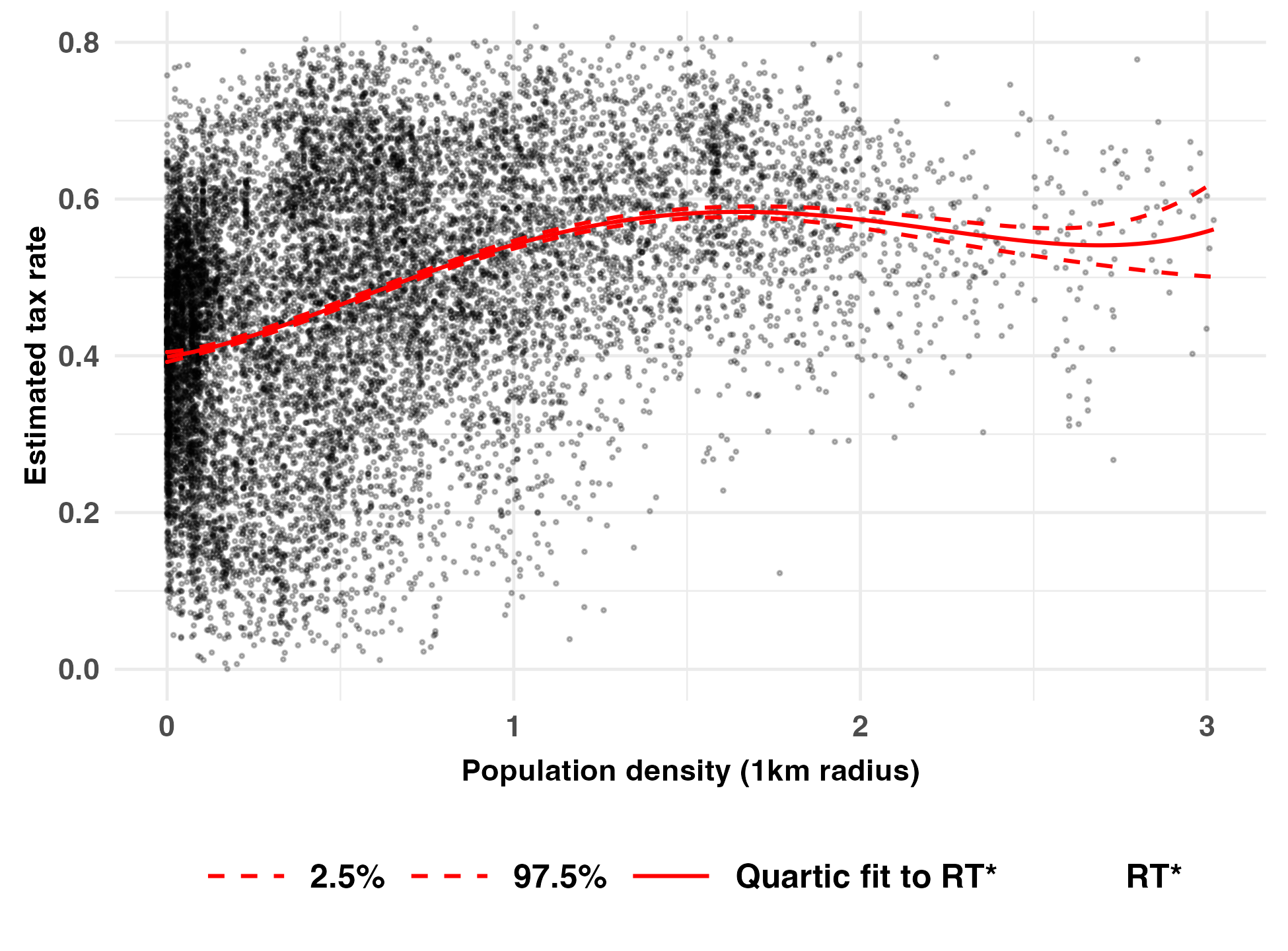}%
				\subcaption{\footnotesize Tax rates by density (1km radius).}
			\end{subfigure}\hfill
			\begin{subfigure}{0.495\linewidth}
				\centering
				\includegraphics[width=\linewidth]{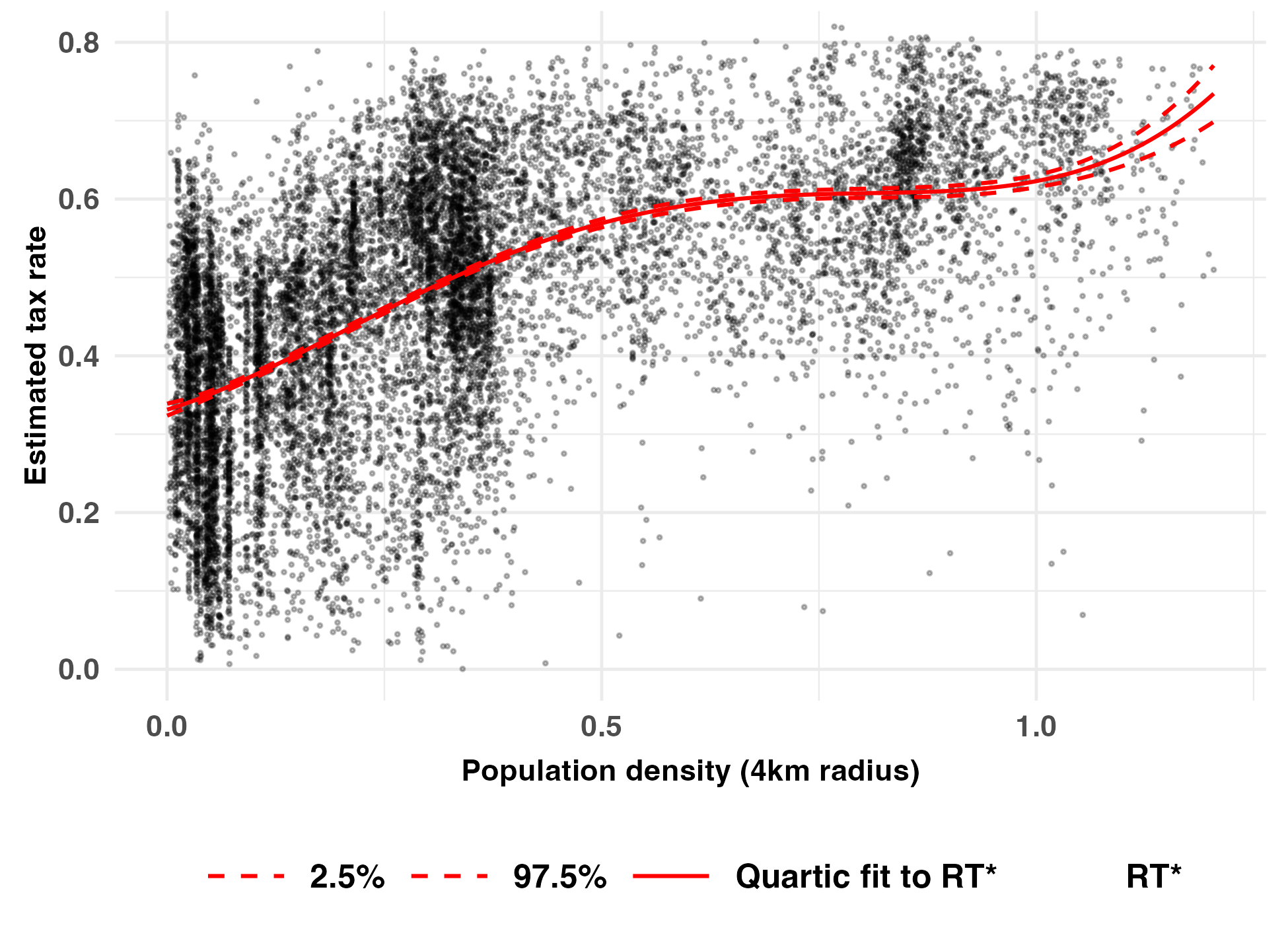}%
				\subcaption{\footnotesize Tax rates by density (4km radius).}
			\end{subfigure}
		}
	\end{adjustwidth}
	
	\caption{\footnotesize (a) Estimated regulatory tax rate by density (10,000's per km\textsuperscript{2}), 1km radius, with quartic fit and 95\% pointwise confidence bands; (b) same for the 4km radius.}
	\label{fig:dens}
\end{figure}

We next examine whether between-locality variation in the regulatory tax rate is correlated with local market conditions. Figure \ref{fig:dens} shows the positive correlation between density and regulatory tax rate. A 4km radius covers over 50 km$^2$, which encompasses nearly the entire built-up area in all but the largest localities (e.g., Jerusalem), and in some cases may extend across municipal boundaries. As such, the 4km density measure can be viewed as a proxy for locality-level density. On average, a 10,000-person increase per km$^2$ is correlated with a 10.3 percentage points higher tax rate using the 1km measure and a 31.5 percentage points increase using the 4km measure. This is confirmed by Regressions (1) and (2), which regress the regulatory tax rate on the two density measures in Table \ref{table:regregs}, excluding locality fixed effects. 
The $R^2$ at the building level is 0.125 for the 1km measure and 0.304 for the 4km measure, but the explanatory power is almost entirely absorbed once locality fixed effects are included as in Regressions (5)–(7).

\begin{figure}[tp]
	\centering
	\includegraphics[width=\linewidth]{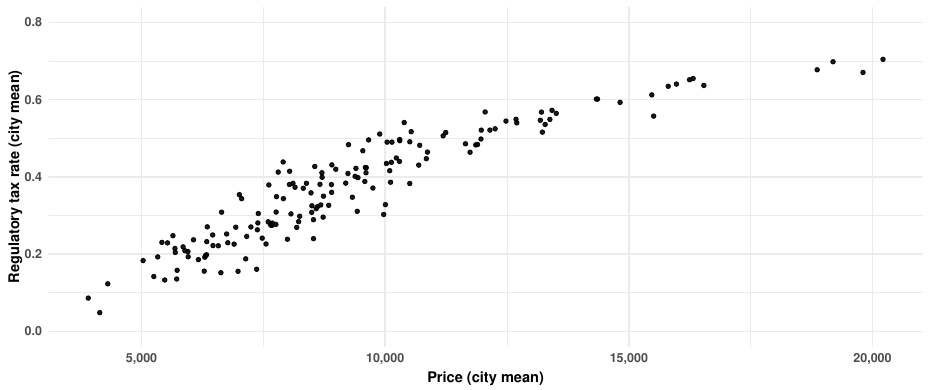}
	\caption{\footnotesize \hspace{-0.3cm}  The locality mean regulatory tax rate against the locality mean apartment price.}
	\label{fig:regpricebycity}
\end{figure}

Figure \ref{fig:regpricebycity} plots the locality mean regulatory tax rate against mean housing prices, showing a tight positive correlation. A simple regression yields an $R^2$ of 0.84, implying that most of the variation in the regulatory tax rate between localities is correlated with local price levels. This is consistent with a flat production frontier and with models that predict greater regulatory restrictiveness in high-amenity, high-demand cities \citep{hilber2013origins}. However, this pattern is not mechanically implied by the estimation: for example, if multi-unit restrictions were concentrated in low-demand suburbs, the correlation could be negative.

\captionsetup{width=\textwidth,justification=raggedright}
\captionsetup[subfigure]{font=footnotesize,justification=raggedright,singlelinecheck=false,skip=2pt}

\captionsetup{width=\textwidth,justification=raggedright}

\FloatBarrier
\subsection{Case Studies: Regulation Over Time in Newly Established Localities} \label{se:newcities}

\begingroup
\setlength{\textfloatsep}{5pt plus 1pt minus 1pt}
\setlength{\floatsep}{5pt plus 1pt minus 1pt}
\captionsetup{width=.94\textwidth,justification=raggedright,font=footnotesize}
\captionsetup[subfigure]{font=footnotesize,justification=raggedright,
	singlelinecheck=false,skip=2pt}

\begin{figure}[b!]
	\centering
	\begin{adjustwidth}{-1.2em}{-1.2em}
		\makebox[\textwidth][c]{%
			\begin{subfigure}[t]{0.34\textwidth}
				\centering
				\includegraphics[width=\linewidth,height=0.21\textheight]{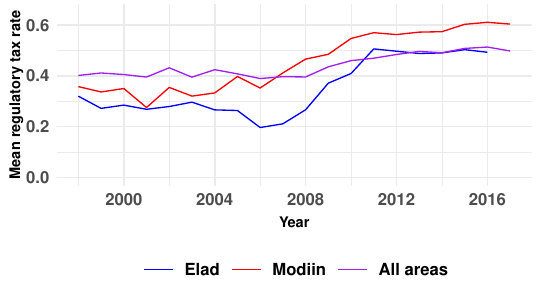}
				\vspace{-0.6em}
				\subcaption{\footnotesize Mean regulatory tax rate.}
				\label{fig:newcitiesregs}
			\end{subfigure}\hspace{0.01\textwidth}%
			\begin{subfigure}[t]{0.34\textwidth}
				\centering
				\includegraphics[width=\linewidth,height=0.21\textheight]{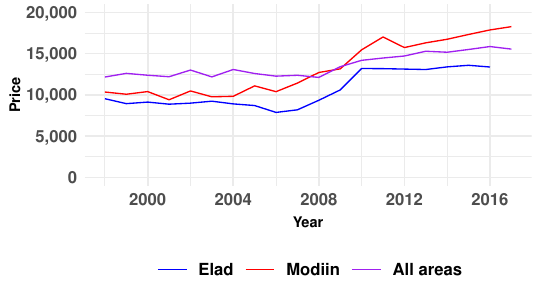}
				\vspace{-0.6em}
				\subcaption{\footnotesize Mean price.}
				\label{fig:newcitiesprices}
			\end{subfigure}\hspace{0.01\textwidth}%
			\begin{subfigure}[t]{0.34\textwidth}
				\centering
				\includegraphics[width=\linewidth,height=0.21\textheight]{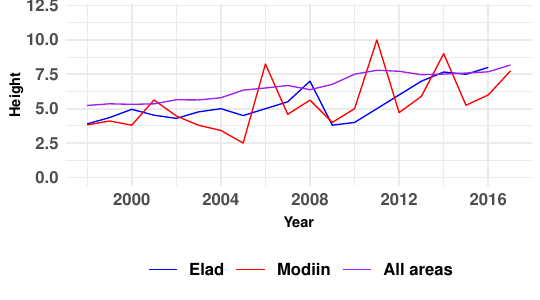}
				\vspace{-0.6em}
				\subcaption{\footnotesize Mean building height.}
				\label{fig:newcitiesheights}
			\end{subfigure}%
		}
	\end{adjustwidth}
	
	\caption{\footnotesize Locality comparisons: regulatory tax, prices, and building heights over time.}
	\label{fig:newcities}
\end{figure}
\endgroup

The newly established localities of Modiin (situated about halfway between Tel Aviv and Jerusalem)
and Elad (about 25 kilometers east of Tel Aviv) offer useful case studies.  
Modiin and Elad were planned in the 1990s. Modiin's first residents arrived in 1996 and Elad's in 1998.  
By 2019, Modiin had about 90,000 residents, most of high socioeconomic status,
while Elad had about 50,000 residents, most religious and of low socioeconomic status.
Since many political economy models of housing regulation locate the source of regulation in home owners' attempts to increase, or at least protect, the asset value of their home, it is useful to document the degree of regulation in newly established localities, before homeowners become politically influential.  

Figure \ref{fig:newcitiesregs} shows the mean estimated regulatory tax rates  for the full sample (in red), in Elad (in purple) from its year of establishment, and in Modiin (in blue) from two years after its establishment (the first year in our data).  
Elad's first residents moved in about two years after Modiin's, and Elad's curve, shifted three years to the left and a few points up, basically overlaps Modiin's curve.
The figure shows that in their nascent years the regulatory tax rates were, although not zero, much lower than the national average, and relatively stable.  
Then about six to eight years after their first residents moved in, the regulatory tax rates essentially doubled.
Modiin's rate settled above the national average, while Elad's at the national average.  
Thereafter, their rates continue to increase at the national rate.
Figures \ref{fig:newcitiesprices} and \ref{fig:newcitiesheights} show that
the increase in regulation is coincident with a jump up in prices yet relatively stable building heights, suggesting that the sudden increase in the regulatory tax was driven by restrictions that were relatively fixed over time, and became more binding with the price increase.

        \section{Conclusion} \label{se:conc}
        
Housing regulation can take many different forms that are often difficult to measure and aggregate, may be arbitrarily enforced,  and is endogenous to building location, market conditions and price. 
Hence, estimating non-land mean costs by a conditional regression embeds unobserved regulatory conditions potentially biasing these estimates. 
In this paper, we show how to identify and estimate frontier costs in multi-floor housing using just observed prices and heights,
identifying frontier marginal costs for heights above MES from variation in demand in unregulated markets and identifying frontier average costs for heights below MES from variation in demand and regulation. 
We adjust prices based on observed apartment floor and building height,
and take into account building-level and apartment-level random housing quality differences and other measurement errors. When quality differs systematically over location and time, we assume  local (weak) complementarity between quality and amenities and bound the regulatory tax.

Using data for newly constructed buildings in the Israeli housing market from 1998-2017, we estimate regulatory tax rates, finding a mean rate of 47\%, with a standard deviation of 17\%.  
Regulatory tax rates are higher in areas that are higher priced, denser, and closer to locality centers.
Measurement errors are small compared to regulation.
When allowing for systematic differences in  quality over location and time, we bound the mean regulatory tax rate in 2017 by 38\% (using buildings within a 2km radius) and 53\%.
Most of that bound is derived from the availability of data on nearby buildings that were built during lower-priced periods and at heights where frontier costs did not significantly decrease. This is contingent on our estimates of a near-zero relationship between temporal demand shocks (period effects) and structural quality (cohort effects). 

There is no presumption that regulation is either welfare-enhancing or welfare-detracting---a determination that would require additional sources of information.  Nor is there a claim that the elimination of all regulation would lead to price reductions in the amount of the estimated regulatory tax, as the resultant price change would require one to know, at a bare minimum, the elasticity of overall housing demand.  Rather the regulatory tax is a measure of the extent of regulation in the market.

Our analysis of regulation is price-based, defining a regulatory tax that relies on vertical deviations from the frontier (i.e., the difference between a building and frontier price at the building height). 
A quantity-based alternative would rely on horizontal deviations from the frontier (i.e., the difference between a building and frontier height at the building price). 
For example, in a counterfactual world where there is no regulation, and holding prices constant, our point estimates indicate that suppliers would build about 5.3 times higher, constructing about 3,454 buildings instead of the 18,171 or so in our sample, and so freeing up about 81\% of the building footprint.
Assessing the resource savings in this counterfactual world would require values for land and consideration of general equilibrium effects, as well as externalities such as congestion effects.  One simple exercise, however, is to consider the land savings from building all apartments in buildings of heights 11 to 24, where marginal costs are constant according to our constrained MLE, in 24-story buildings instead.  
This would require 36\% less land, but cost an additional 2\% of non-land costs.
Likewise, removing regulation so that apartments in shorter than MES-story buildings are built in MES buildings would require 38\% less land, along with saving 4\% of non-land costs. 
We leave further analysis along these lines for future work.	 

\begingroup
\setstretch{1.12}
\bibliography{biblio}
\endgroup

\pagebreak

\appendix

\noindent {\LARGE \bf Online Appendix}

\section{Additional Estimation Details}\label{ap:est}

\subsection{Apartment-Floor, Building-Height Adjusted Prices} \label{ap:PSR}

This section expands on the brief discussion of Subsection \ref{se:preresults} to provide additional details on construction of the adjusted prices used in the frontier estimation and regulatory tax estimates of Section \ref{se:results}.  
Using the dataset of new apartment transactions described in Subsection  \ref{se:data}, we conducted a preliminary analysis by regressing the log of the real, cost adjusted, per square meter price on a full set of floor and building height interactions, dummy variables for transaction year before and transaction year after the year of construction, a nine-degree polynomial in the calendar day of transaction, eight dummies for the legal status of the property,
and dummy variables for the land parcel. 
Identification of the floor effects is possible because of cases in which there are transactions of multiple apartments in the same building, but on different floors.  Identification of the height effects is possible because of cases in which there are multiple buildings on the same land parcel.\footnote{These are a small fraction of the data, but of sufficient number that the height effects can be measured.  The frontier estimation is conducted on a sample in which apartments in the 534 buildings sitting on these land parcels are removed.}

\begin{figure}[tp]
	\centering
	\begin{adjustwidth}{-1.2em}{-1.2em}
		\makebox[\linewidth][c]{%
			\begin{subfigure}{0.495\linewidth}
				\centering
				\includegraphics[width=\linewidth]{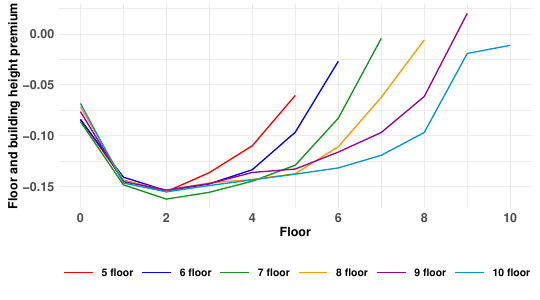}%
				\subcaption{\footnotesize 5--10 floor and building height effects.}
				\label{fig:PSR1}
			\end{subfigure}\hfill
			\begin{subfigure}{0.495\linewidth}
				\centering
				\includegraphics[width=\linewidth]{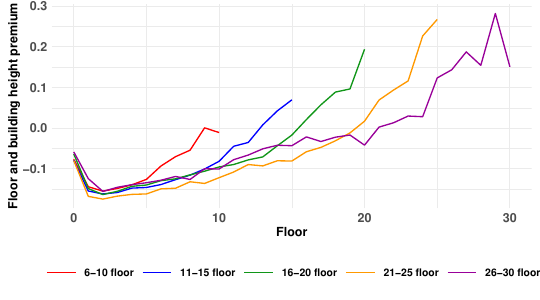}%
				\subcaption{\footnotesize 5--30 floor and building height effects.}
				\label{fig:PSR2}
			\end{subfigure}
		}
	\end{adjustwidth}
	\caption{\footnotesize Floor and building height effects.}
	\label{fig:PSR}
\end{figure}

A selected set of the estimates for the floor $\times$ height interactions in buildings with 5 to 10 floors are shown in Figure \ref{fig:PSR1}.  For given building height, the relationship between price and floor is J-shaped and right-leaning, with price falling initially, reflecting an initial preference for the ground floor and then more or less linearly increasing, until a penthouse effect at the penultimate and top floor.  
There is also a building height effect, with shorter buildings preferred to taller ones, especially at higher floors.  Figure \ref{fig:PSR2} covers a wider range of heights, grouping each 5 floor range of heights, and shows similar results.

On the basis of these estimates, we choose to model the conditioning on floor and height by a linear term in floor, dummy variables for each of the ground, first, second, and third floors, a linear term in building height, and dummies for the penultimate and top floors, as well as interaction with the sum of those two dummies and the building height.  There are also interactions between a dummy for above four floors with the first, second, and third floor dummies, and interactions between heights above 10 floors and the linear term in floor.\footnote{These two cutoffs originate in the minimal regulatory requirements for a first and a second elevator.}  Table \ref{table:PSR} presents the coefficients and standard errors of the main variables.  

\captionsetup{width=\textwidth,justification=raggedright}
\captionsetup[subfigure]{font=footnotesize,justification=raggedright,singlelinecheck=false,skip=2pt}

\begin{table}[!htb]
	\centering
	\begin{threeparttable}
		\caption{Preliminary stage regression}
		\label{table:PSR}
		\small
		\begin{tabular}{@{}lc@{}}
			\toprule
			& Log price \\
			\midrule
Floor                             & \makecell{ 0.0088 \\ \footnotesize (0.0003)} \\
Building height                   & \makecell{ -0.0006 \\ \footnotesize (0.0001)} \\
Penthouse                         & \makecell{ 0.0361 \\ \footnotesize (0.0016)} \\
Penthouse $-1$                    & \makecell{ 0.0058 \\ \footnotesize (0.0017)} \\
Penthouse $\times$ building height& \makecell{ 0.0027 \\ \footnotesize (0.0002)} \\
Year before construction year     & \makecell{ -0.0037 \\ \footnotesize (0.0009)} \\
Year after construction year      & \makecell{ 0.0030 \\ \footnotesize (0.0007)} \\
\bottomrule

		\end{tabular}
	\end{threeparttable}
	
			\vspace{4pt}
	\begin{minipage}{\textwidth}
		\footnotesize
		\textit{Notes:} Standard errors are in parentheses.
		Additional controls: polynomial in calendar time; dummies for ground, first, second, and third floors and their interactions with dummies for building heights above 4 and 10 floors; eight legal status dummies; and parcel fixed effects.
	\end{minipage}
\end{table}

	\subsection{Estimating $\kappa_T$} \label{ap:timeeffects}

Our aim is to estimate $\kappa_T$ through the relationship between period effects (transaction time) and cohort effects (construction time) in a regression of existing home prices on period, cohort, and age (transaction time less construction time, capturing depreciation), where the cohort effects are restricted to be a function of the period effects.  In its most general form, this entails estimating 
\begin{align*}
	y_{its} &=\gamma (t) + \delta (\gamma (s)) + \alpha (t-s),
\end{align*}
where $s$ is construction period, $t$ is transaction period (so that $t-s$ is age), $\gamma(t)$ (which corresponds to its namesake in Subsection \ref{se:estRT}) are period effects, $\delta (\gamma(s))$ are cohort effects, and $\alpha(t)$ are age effects.  This restriction on the cohort effects is implied by the model outlined in Subsection \ref{se:estRT}, where cohort effects capture variations in housing quality over time. So long as $\gamma$ is nonlinear, the restriction provides one solution to the well-known problem of decomposing a variable into age, period, and cohort effect, as period is the sum of cohort and age \citep[e.g.,][]{hall20138,hall2007identifying}.  A number of different approaches have been taken in the hedonic pricing literature  \citep[e.g.,][]{coulson2008estimating}.  Our approach is dictated by our goal of estimating $\kappa_T$ and the theoretical framework in Subsection \ref{se:estRT} which motivates that objective.

We set $\gamma$ and $\alpha$ to be quadratic functions, and, as we are after only a single number for $\kappa_T$, set $\delta$ as a constant.  Nonlinearity is essential, as $\delta$ is unidentified if $\gamma$ is linear.  Thus we estimate,
\begin{align*}
	y_{its} &=\gamma _{1}t+\gamma _{2}t^{2}+\delta (\gamma _{1}s+\gamma
	_{2}s^{2})+\alpha _{1}(t-s)+\alpha _{2}(t-s)^{2}.
\end{align*}

A consistent estimate for $\delta$ can be obtained by regressing log price on the period of transaction and its square, the square of the period of construction, age (or period of construction) and age-squared. The estimate $\widehat \delta$   is the ratio of the coefficient on the square of the period of construction to the coefficient on the square of the period of transaction.  Column (1) in Table \ref{tab:EHPR} shows the results of the regression, with parcel fixed effects and the same set of building and apartment attributes as in Table \ref{table:PSR} of Appendix \ref{ap:PSR}.  The regression is estimated on the data described in Section \ref{se:data} but expanding the dataset to include transactions of not only new but also existing apartments, specifically all transactions with construction years the year after or up to 40 years before the transaction year.

We estimate $\widehat \delta = 0.0005/0.311 = 0.0016$ ($s.e. = 0.0018$), and so $\widehat \kappa_T = \widehat \delta / (1+\widehat \delta) = 0.0016$ (s.e. = 0.0018), indicating that housing quality barely varies with price over time. We obtain similar results for $\gamma$ and $\alpha$ quartic functions.

Column (2) in Table \ref{tab:EHPR} drops the squared year of construction, substituting instead its interaction with indicator functions for the twenty largest (by number of transactions) localities and an indicator for all other localities.  This allows the relationship between period effects and cohort effects to vary across locations. The results are very similar.  No locality shows an absolute ratio exceeding $0.0460$, while the ratio of the weighted mean of the interaction coefficients to the square of the transaction year (with weights equal to the frequency of the localities and the residual category in the regression sample) is  $-0.0037$ (s.e. =  $0.0019$).

\begin{table}[!htb]
	\centering
	\begin{threeparttable}
		\caption{Existing homes price regression}
		\label{tab:EHPR}
		\small
		\begin{tabular}{@{}lcc@{}}
			\toprule
			Variable & (1) & (2) \\
			\midrule
Year of transaction              & -0.034  & -0.033  \\
& (0.001) & (0.001) \\
Year of transaction squared/100  &  0.311  &  0.310  \\
& (0.002) & (0.002) \\
Year of construction squared/100 &  0.0005 & --      \\
& (0.001) &         \\
Age                              &  0.0012 &  0.0012 \\
& (0.0002)& (0.0002)\\
Age squared/100                  & -0.0036 & -0.0033 \\
& (0.0007)& (0.0007)\\
\bottomrule

		\end{tabular}
	\end{threeparttable}
	
				\vspace{4pt}
	\begin{minipage}{\textwidth}
		\footnotesize
		\textit{Notes:} The dependent variable is log price per square meter in real 2017 NIS. 
		Year is calendar year minus 1997. The number of observations is 776,709.
	\end{minipage}
\end{table}

\FloatBarrier			
\subsection{Variances}\label{ap:var}

Conditioning  on height, we estimate the variances of $u$, $v$, and $w$ using  apartment, building, and bloc hierarchical modeling,
\begin{align}
	\widehat \Var(v) & =\frac{1}{\sum_{k=1}^K\sum_{i=1}^{n_k}(J_{ki}-1)}\sum_{k=1}^K\sum_{i=1}^{n_k}\sum_{j=1}^{J_{ki}}(y_{kij}^0-\bar y_{ki}^0)^2, \label{varv2}\\
	\widehat \Var(w)&=\frac{1}{\sum_{k=1}^K(n_k-1)} \Big(\sum_{k=1}^K\sum_{i=1}^{n_k}(\bar y_{ki}^0-\bar y_{k}^0)^2	-  	\widehat \Var(v)\sum_{k=1}^K\sum_{i=1}^{n_k}\frac{n_k-1}{n_kJ_{ki}}\Big),\label{varw2}\\
	\widehat \Var(u)&=
	\frac{1}{K-1}\sum_{k=1}^K\Big(\bar y_k - \bar{y}\Big)^2	
	- \frac{	\widehat \Var(w)}{K}\sum_{k=1}^K\frac{1}{n_{k}} 
	-  \frac{	\widehat \Var(v)}{K}\sum_{k=1}^K \sum_{i=1}^{n_k}\frac{1}{n_{k}^2J_{ki}},  \label{varu2}
\end{align}
where $y_{kij}^0$ is the residual of a nonparametric series regression of log price on transaction date (in days), 
and where the estimated building prices are
$\bar y^0_{ki} =\frac{1}{J_{ki}}\sum_{j=1}^{J_{ki}}y^0_{kij}$, $\bar y_{ki} =\frac{1}{J_{ki}}\sum_{j=1}^{J_{ki}}y_{kij}$,
the estimated bloc prices are
$\bar y^0_{k} =\frac{1}{n_k}\sum_{i=1}^{n_k}\bar y^0_{ki}$ and  $\bar y_{k} =\frac{1}{n_k}\sum_{i=1}^{n_k}\bar y_{ki}$,
and the overall average prices are
$\bar y^0 =\frac{1}{K}\sum_{k=1}^K\bar y_{k}^0$ and $\bar y =\frac{1}{K}\sum_{k=1}^K\bar y_{k}$.

\subsection{The Frontier}\label{ap:frontest}

Fix height $h$. To simplify notation, drop the height index $h$.
Since $u\sim TN(\mu_u,\sigma_u^2)$,
\begin{align}
	\Var(u)&= \sigma_u^2\Big[1- \frac{\mu_u}{\sigma_u}\cdot \lambda\Big(\frac{ \mu_u}{ \sigma_u}\Big)
	-\Big(\lambda\Big(\frac{ \mu_u}{ \sigma_u}\Big)\Big)^2 \Big] ,\label{eq:varu}
\end{align}
where $\lambda(x)= \phi(x)/\Phi(x) $, and $\phi(.)$ and $\Phi(.)$ are the standard normal probability and cumulative density functions.
Combining  \eqref{varu2} with \eqref{eq:varu} 
we obtain,
{\small\begin{align}
		\widehat \sigma_u^2\Big[1- \frac{\widehat\mu_u}{\widehat\sigma_u}\cdot \lambda\Big(\frac{\widehat \mu_u}{ \widehat\sigma_u}\Big)
		-\Big(\lambda\Big(\frac{ \widehat  \mu_u}{ \widehat \sigma_u}\Big)\Big)^2 \Big]
		&=\frac{1}{K-1}	\sum_{k=1}^K\big(\bar y_k - \bar{y}\big)^2	
		- \frac{	\widehat \sigma_w^2}{K}\sum_{k=1}^K\frac{1}{n_{k}} 
		-  \frac{	\widehat \sigma_v^2}{K}\sum_{k=1}^K \sum_{i=1}^{n_k}\frac{1}{n_{k}^2J_{ki}}.  \label{gmuusigu2}
	\end{align}
}So that given the data and parameters $\widehat\mu_u$,  $	\widehat \sigma_v^2$, and $	\widehat \sigma_w^2$,
we obtain $\widehat \sigma_u^2$ using \eqref{gmuusigu2}.

For each of $M$ parameter values for $(g,\mu_u)$ 
and the estimates for $\sigma_v^2$ and $ \sigma_w^2$ from \eqref{varv2}-\eqref{varu2} we obtain an estimate for $\sigma_u^2$ and calculate the log likelihood (ignoring constants),
\begin{align}
	&	\mathcal{L}_h(g,\mu_{u},\sigma_{u}^2,\sigma_v^2,\sigma_w^2;\cdot)	=\frac{1}{2}\sum_{k=1}^K \Big(	\frac{\mu_k^2}{\sigma_k^2}-\frac{\mu_u^2}{\sigma_u^2}+\frac{1}{\sigma_v^2}\sum_{i=1}^{n_k}\Big(\frac{\sigma_w^2(\sum_{j=1}^{J_{ki}}(y_{kij}-g))^2}{\sigma_v^2 +J_{ki}\sigma_w^2}-\sum_{j=1}^{J_{ki}}(y_{kij}-g)^2\Big) +
	\nonumber	\\ 
	&\ln \sigma_k^2 	-\ln  \sigma_u^2-\sum_{i=1}^{n_k} \Big(\ln  (\sigma_v^2 +J_{ki}\sigma_w^2)	
	+ (J_{ki}-1)\ln  \sigma_v^2 \Big)
	+2\ln \Phi\big(\frac{\mu_{k}}{\sigma_{k}}\big) -2\ln  \Phi \big(\frac{\mu_u}{\sigma_u}\big)\Big), \label{loglik}\\
		&	\mu_{k}=\frac{\sigma_{k}^2}{\sigma_u^2n_k} \sum_{i=1}^{n_k}\frac{\mu_u(\sigma_v^2 +J_{ki}\sigma_w^2)+n_k\sigma_u^2\sum_{j=1}^{J_{ki}}(y_{kij}-g) }{\sigma_v^2 +J_{ki}\sigma_w^2} , \nonumber \\
	&\sigma_{k}^2=\sigma_u^2n_k\Big(\sum_{i=1}^{n_k}\frac{\sigma_v^2 +J_{ki}\sigma_w^2+n_kJ_{ki}\sigma_u^2 }{\sigma_v^2 +J_{ki}\sigma_w^2}\Big)^{-1}, 	 \nonumber
\end{align}
where 
$\mu_{k}$ is a weighted average of $\mu_u$ and the average distance of log price to the frontier. 

Now, the global maximum of the likelihood at height $h$ is obtained by maximizing \eqref{loglik}.
The global maximum of the likelihood, constrained so that average costs decrease to MES and marginal costs increase thereafter, is attained by a grid search and Dijkstra's algorithm,
\begin{align*}
	&	\{\widehat{MES},\widehat g,\widehat \mu_u\} =\argmax_{
		\substack{mes \in \{1,\ldots,H-1\}   \\
			\textsl{g}\in \mathbb{R}^H,
			\nu_u \in \mathbb{R}^H}}  
	\	\sum_{h=1}^H\mathcal{L}_h(\textsl{g}_h,\nu_{uh},\cdot )  ,\\
	&\text{s.t. }  \textsl{g}_{mes} \leq \textsl{g}_{mes-1} \leq \ldots \leq \textsl{g}_1
	\textrm{ and }	\textsl{g}_{mes} \leq \textsl{g}_{mes+1} \leq \ldots \leq \textsl{g}_H.  
\end{align*}

We now derive the likelihood in \eqref{loglik}.
Assume $v_{kij}\sim N(0,\sigma_v^2)$, $w_{ki}\sim N(0,\sigma_w^2)$, and $u_{k}\sim TN(\mu_u,\sigma_u^2)$. So,
\begin{align*}
	f_{ v_{kij}}(v) &= \frac{e^{-v^2/2\sigma_v^2}}{\sqrt{2\pi \sigma_v^2}}, &&
	f_{w_{ki}}(w) = \frac{e^{-w^2/2\sigma_w^2}}{\sqrt{2\pi \sigma_w^2}}, &&
	f_{u_{k}}(u) = \frac{e^{-(u-\mu_u)^2/2\sigma_u^2}}{\sqrt{2\pi \sigma_u^2} \cdot \Phi(\mu_u /\sigma_u)}, & & u \geq 0 .
\end{align*} 
By independence of $u_{k}, w_{k1},\ldots,w_{kn_k},v_{k11},\ldots,v_{k1J_{k1}},\ldots, v_{kn_k1},\ldots,v_{kn_kJ_{kn_k}}$, 
\begin{align*}    
	&f_{u_{k}+w_{k1}+v_{k11},\ldots,u_{k}+w_{k1}+v_{k1J_{k1}},\ldots,u_{k}+w_{kn_k}+v_{kn_k1},\ldots,u_{k}+w_{kn_k}+v_{kn_kJ_{kn_k}}}(s_{11},\ldots,s_{1J_{k1}},\ldots,s_{n_k1},\ldots,s_{n_kJ_{kn_k}}) \\
	&=\int_{0}^{\infty}\int_{-\infty}^{\infty}\cdots \int_{-\infty}^{\infty} f_{u_{k}}(u)  
	\prod_{i=1}^{n_k} \Big(f_{w_{ki}}(w_i) \prod_{j=1}^{J_{ki}}  f_{v_{kij}}(s_{ij}-w_i-u)  dw_i\Big) du\\
	&=\int_0^{\infty}\frac{e^{-(u-\mu_u)^2/2\sigma_u^2}}{\sqrt{2\pi \sigma_u^2} \cdot \Phi(\mu_u/\sigma_u)} 
	\prod_{i=1}^{n_k}	\Big( \int_{-\infty}^{\infty} \frac{e^{-w_i^2/2\sigma_w^2}}{\sqrt{2\pi \sigma_{w}^2}} \prod_{j=1}^{J_{ki}} \frac{e^{-(s_{ij}-w_i-u)^2/2\sigma_v^2}}{\sqrt{2\pi \sigma_v^2}} dw_i\Big)du\\
	&=\frac{\sigma_k\exp\Big(\sum_{i=1}^{n_k}\frac{\sigma_w^2(\sum_{j=1}^{J_{ki}}s_{ij})^2}{2(\sigma_v^2 +J_{ki}\sigma_w^2)\sigma_v^2}-\frac{\mu_u^2}{2\sigma_u^2}-\sum_{i=1}^{n_k}\frac{\sum_{j=1}^{J_{ki}}s_{ij}^2}{2\sigma_v^2}	+\frac{\mu_k^2}{2\sigma_k^2}\Big)\Phi({\mu_{k}}/{\sigma_{k}})}
	{(2\pi)^{\frac{1}{2}\sum_{i=1}^{n_k}J_{ki}} \sigma_u \Phi(\mu_u/\sigma_u) \sigma_v^{\sum_{i=1}^{n_k} (J_{ki}-1)}\prod_{i=1}^{n_k}\sqrt{\sigma_v^2 +J_{ki}\sigma_w^2} },
\end{align*}
where
\begin{align*}
		\mu_{k}&=\frac{\sigma_{k}^2}{\sigma_u^2n_k} \Big(\sum_{i=1}^{n_k}\frac{\mu_u(\sigma_v^2 +J_{ki}\sigma_w^2)+n_k\sigma_u^2\sum_{j=1}^{J_{ki}}s_{ij} }{\sigma_v^2 +J_{ki}\sigma_w^2}\Big),\\
	\sigma_{k}^{2}&=\sigma_u^2n_k\Big(\sum_{i=1}^{n_k}\frac{\sigma_v^2 +J_{ki}\sigma_w^2+n_kJ_{ki}\sigma_u^2 }{\sigma_v^2 +J_{ki}\sigma_w^2}\Big)^{-1}.
\end{align*}
We show $u|u+\eta$ is truncated normal in \eqref{eq:distu1}. 
Assume $u \sim TN(\mu_u,\sigma_u^2)$ and $\eta \sim N(0,\sigma_\eta^2) $. 
{\begin{align*}    
		f_{u,u+\eta}(u,s)
		&=\frac{e^{-(u-\mu_u)^2/2\sigma_u^2}e^{-(s-u)^2/2\sigma_\eta^2}}{2\pi \sigma_u\sigma_\eta \cdot \Phi(\mu_u/\sigma_u)} ,\\
		f_{u+\eta}(s)&=\int_{0}^{\infty} f_{u}(u) f_{\eta}(s-u) du
		=\frac{\sigma_* \exp\big(\frac{\mu_*^2}{2\sigma_*^2} -\frac{\mu_u^2}{2\sigma_u^2} -\frac{s^2}{2\sigma_\eta^2}  \big)\Phi(\mu_*/\sigma_*)}{\sqrt{2\pi} \sigma_u  \sigma_{\eta}\cdot \Phi(\mu_u/\sigma_u)} ,\\
		f_{u|u+\eta}(u|s)
		&= \frac{\exp\big(-\frac{(u-\mu_u)^2}{2\sigma_u^2} -\frac{(s-u)^2}{2\sigma_\eta^2}-\frac{\mu_*^2}{2\sigma_*^2} +\frac{\mu_u^2}{2\sigma_u^2} +\frac{s^2}{2\sigma_\eta^2}\big)}{\sqrt{2\pi}\sigma_* \Phi(\mu_*/\sigma_*) }   
		= \frac{\exp\big(-\frac{1}{2\sigma_*^2}(u-\mu_*)^2\big)}{\sqrt{2\pi}\sigma_* \Phi(\mu_*/\sigma_*) } , 
	\end{align*} 
}where
$	\mu_{*}=\dfrac{\sigma_\eta^2\mu_u+s\sigma_u^2}{\sigma_u^2+\sigma_\eta^2}$ and  
$\sigma_{*}^{2}= \dfrac{\sigma_u^2\sigma_\eta^2}{\sigma_u^2+\sigma_\eta^2}$.

\FloatBarrier
\section{The Frontier Isoquant and Elasticity of Substitution of Land for Capital} \label{ap:eos}

The isoquant in Figure \ref{fig:eosiso} is constructed from the discrete frontier cost schedule. For each height $h$, the plotted point is $(1/Q_h,C_h/Q_h)$, where $C_h$ is the total non-land cost implied by the frontier. The elasticity of substitution of the housing production function is the rate at which the cost-minimizing capital to land ratio varies with the marginal rate of technical substitution. This is commonly used to summarize the degree of substitution of one input for the other in housing production.  

Given price taking firms in the input market, and normalizing the price of capital to 1, the elasticity of substitution is,
\begin{align*}
	\sigma &=\frac{d\ln k}{d\ln R}=\frac{R}{k} \times \frac{dk}{dR},   
\end{align*}
where $k=K/L$ is the capital to land ratio (or the capital per  {unit of land}), $K$ is capital, $L$ is a given fixed amount of land, 
and $R$ is the price of one {unit of land}, i.e., land rent. 

With the constant returns to scale production function in land and capital $f_0(K,L)$, per  {unit of land} housing output, equivalently height $h$, satisfies $h=f_{0}(K,L)/L=f_{0}({K}/{L},1)=f(k).$
Noting that $k=C(h)$, $h=C^{-1}(k)=f(k)$, $C'(h)=1/f'(k)$, and $C''(h)=-f''(k)/(f'(k))^3$,  the elasticity of substitution is,
\begin{align*}
	\sigma  &= \frac{f'(k)(kf'(k)-f(k))}{kf(k)f''(k)}= \frac{C'(h)(hC'(h)-C(h))}{hC(h)C''(h)} =\overbrace{\underbrace{\frac{(MC-AC)\times h}{h \times AC }}_k}^{R}\times \overbrace{\underbrace{\frac{MC\times dh}{h \times dMC}}_{dR}}^{dk} 
	=\frac{d\ln AC}{d \ln MC}, 
\end{align*}
where the first equality follows from \cite{arrow1961capital}.

Since in an unregulated market, housing price equals marginal non-land cost, this is also the elasticity of average non-land cost to market price. Furthermore, since price equals total average cost (the long run, zero profit condition), the elasticity of substitution relates the growth of land rent to the growth of non-land costs as height increases. We report the isoquant rather than a smooth graph of $\sigma$, because $\sigma$ depends on the second derivative of the cost frontier. The data identify the frontier at discrete building heights, but a smooth elasticity requires curvature between heights, which would depend on an additional smoothing choice.

\section{Additional Figures and Tables} \label{ap:figstabs}

\subsection{Local Concentration} \label{ap:comp}
As noted in the main part of the paper, the construction industry in Israel is structurally competitive, with a ten-firm national concentration ratio of 0.15 only \citep{MinFin2017}. That the largest firms are known to operate throughout the country, and the country is geographically small, suggests low local concentration as well.
Lacking information on the builder's identity in the transaction dataset that is the main source for our analysis, we turn to auctions for construction rights on government owned land for a quantitative statement on local concentration.  
The auctions, held from 1998 to 2017 for the approximately fifty percent of construction that takes place on government owned land \citep{rubin2017supply}, can not be reliably matched to our transaction data as their finest, reliable geographical identifier in those data is at the locality level.  Yet they can provide us with a sense of local competition.  We calculate a mean locality-level HHI over all zoned-for dwelling units in the auctions of 0.025, equivalent to forty equally-sized firms.

As a rule, larger markets support more competitors and so are characterized by smaller markups \citep{sutton1991sunk} and less firm heterogeneity, as the more inefficient firms are priced out of the market \citep{syverson2004market}.  This suggests that any bias in the measured regulatory tax arising from varying markups or differential firm efficiency decreases with market size.  Thus the extent of any contribution of varying markups and differential firm efficiency to the measured regulatory tax can be gauged by how regulatory tax varies empirically with measures of market size.  We consider two measures.  The first is population density, used as a proxy for the stock of existing homes in the vicinity of the given apartment or building. Existing homes, whether renovated or not, are substitutes for newly constructed housing, and so both limit markup and, pushing down price, limit inefficient firms.  The second measure is the number of buildings constructed in the vicinity of the apartment or building over the period of our sample. 

Table \ref{table:regcomp} shows how the regulatory tax varies with both population density and new building construction.  As in our previous results, the regulatory tax rate is always positively associated with population density and so with the stock of existing homes.  Furthermore, the regulatory tax rate is also positively associated with new building construction within a 1km radius.  Only when we extend the radius to 4km, which in most cases is large enough to encompass a locality, and condition on fixed effect and distance to the locality center, do we find a negative relationship between construction and the regulatory tax rate. Even in that case, the coefficient is very small:  1000 more buildings is associated with three percentage points lower regulatory tax.  One thousand buildings is essentially the ninetieth percentile of constructed buildings within a 4km radius. The maximum is 1612, which would be associated with a five percentage point decrease. Such numbers are small with respect to the mean regulatory tax rate of 47.4\% that we estimate.

\begin{table}[!htb]
	\centering
	\begin{threeparttable}
		\caption{Regressions}
		\label{table:regcomp}
		\small
\begin{tabular}{@{}lcccc@{}}
\toprule
\multicolumn{5}{@{}l}{\textit{Apartment}} \\
\midrule
\T New building construction (1km radius) & \makecell {\ \ 0.00033 \\ \footnotesize{(0.00004)}} & \ \ - & \makecell {\ \ 0.00003 \\ \footnotesize{(0.00005)}} & \ \ - \\
\T New building construction (4km radius) & \ \ - & \makecell {\ \ 0.00009 \\ \footnotesize{(0.00001)}} & \ \ - & \makecell {\ \ -0.00006 \\ \footnotesize{(0.00002)}} \\
\T Distance to locality center & \ \ - & \ \ - & \makecell {\ \ -0.00378 \\ \footnotesize{(0.00157)}} & \makecell {\ \ -0.00504 \\ \footnotesize{(0.00150)}} \\
\T Density (1km radius) & \makecell {\ \ 0.06794 \\ \footnotesize{(0.00529)}} & \ \ - & \makecell {\ \ 0.00746 \\ \footnotesize{(0.00537)}} & \ \ - \\
\T Density (4km radius) & \ \ - & \makecell {\ \ 0.18456 \\ \footnotesize{(0.01175)}} & \ \ - & \makecell {\ \ 0.11010 \\ \footnotesize{(0.02104)}} \\
\T Locality fixed effects & No & No & Yes & Yes \\
\T $R^2$ & \ \ 0.096 & \ \ 0.246 & \ \ 0.564 & \ \ 0.566 \\
\addlinespace
\multicolumn{5}{@{}l}{\textit{Building}} \\
\midrule
\T New building construction (1km radius) & \makecell {\ \ 0.00017 \\ \footnotesize{(0.00003)}} & \ \ - & \makecell {\ \ 0.00009 \\ \footnotesize{(0.00002)}} & \ \ - \\
\T New building construction (4km radius) & \ \ - & \makecell {\ \ 0.00008 \\ \footnotesize{(0.00000)}} & \ \ - & \makecell {\ \ -0.00003 \\ \footnotesize{(0.00001)}} \\
\T Distance to locality center & \ \ - & \ \ - & \makecell {\ \ -0.00589 \\ \footnotesize{(0.00083)}} & \makecell {\ \ -0.00791 \\ \footnotesize{(0.00082)}} \\
\T Density (1km radius) & \makecell {\ \ 0.09149 \\ \footnotesize{(0.00294)}} & \ \ - & \makecell {\ \ 0.00615 \\ \footnotesize{(0.00301)}} & \ \ - \\
\T Density (4km radius) & \ \ - & \makecell {\ \ 0.23626 \\ \footnotesize{(0.00615)}} & \ \ - & \makecell {\ \ 0.11816 \\ \footnotesize{(0.01287)}} \\
\T Locality fixed effects & No & No & Yes & Yes \\
\T $R^2$ & \ \ 0.129 & \ \ 0.318 & \ \ 0.671 & \ \ 0.673 \\
\bottomrule
\end{tabular}

	\end{threeparttable}
	
					\vspace{4pt}
	\begin{minipage}{\textwidth}
		\footnotesize
		\textit{Notes:} Standard errors are in parentheses underneath the coefficients. 
		Distance to locality center is in kilometers.
		Densities are in 10,000's per square kilometer. 
		There are 13,102 buildings and 206,822 apartments. 
		The top panel has outcomes at the apartment level, with standard errors clustered at the building level.
		The bottom panel has outcomes equal to the average apartment price in a building.
	\end{minipage}

\end{table}

\clearpage
\FloatBarrier
\subsection{Maximum likelihood estimates} \label{ap:mlenum}

\vspace{-0.4cm}
\begin{table}[!htb]
	\centering
	\begin{threeparttable}
		\caption{Maximum likelihood estimates}
		\label{tableMLE}
		\small
		\begin{tabular}{@{}r r r r r r@{}}
			\toprule
			Height & Quantity & MLE & MLE by height & Minimum & Mean \\
			\midrule
1 & 1.05 & 7,411 & 7,411 & 5,666 & 10,544 \\
2 & 2.07 & 6,869 & 6,869 & 5,052 & 11,538 \\
3 & 3.09 & 6,835 & 6,835 & 5,354 & 12,458 \\
4 & 4.09 & 6,730 & 6,730 & 5,385 & 12,757 \\
5 & 5.03 & 6,514 & 6,514 & 5,374 & 14,288 \\
6 & 6.05 & 6,624 & 6,624 & 5,256 & 14,684 \\
7 & 7.07 & 6,729 & 6,787 & 5,842 & 14,347 \\
8 & 8.10 & 6,729 & 6,858 & 5,319 & 14,069 \\
9 & 9.14 & 6,729 & 6,697 & 5,705 & 14,007 \\
10 & 10.19 & 6,729 & 6,511 & 5,605 & 14,277 \\
11 & 11.18 & 7,131 & 7,367 & 6,576 & 15,555 \\
12 & 12.23 & 7,131 & 6,983 & 6,777 & 15,839 \\
13 & 13.28 & 7,131 & 7,614 & 7,501 & 15,498 \\
14 & 14.35 & 7,131 & 6,232 & 6,078 & 14,000 \\
15 & 15.42 & 7,131 & 7,813 & 6,503 & 14,568 \\
16 & 16.50 & 7,131 & 6,199 & 6,410 & 15,252 \\
17 & 17.58 & 7,131 & 6,974 & 7,103 & 15,491 \\
18 & 18.68 & 7,131 & 8,029 & 7,508 & 15,150 \\
19 & 19.78 & 7,131 & 6,769 & 6,940 & 15,228 \\
20 & 20.89 & 7,131 & 6,770 & 7,156 & 15,221 \\
21 & 22.00 & 7,131 & 8,831 & 8,901 & 16,847 \\
22 & 23.13 & 7,131 & 7,715 & 8,753 & 16,515 \\
23 & 24.26 & 7,131 & 10,591 & 8,919 & 15,569 \\
24 & 25.40 & 7,131 & 6,711 & 7,433 & 18,155 \\
25 & 26.54 & 8,253 & 9,621 & 9,591 & 16,903 \\
26 & 27.69 & 8,253 & 11,093 & 11,015 & 14,778 \\
27 & 28.86 & 8,253 & 12,404 & 12,820 & 19,637 \\
28 & 30.03 & 8,253 & 7,901 & 8,479 & 18,088 \\
29 & 31.20 & 9,203 & 9,741 & 10,157 & 19,635 \\
30 & 32.39 & 9,203 & 8,772 & 9,637 & 21,399 \\
31 & 33.58 & 9,632 & 9,632 & 10,742 & 24,481 \\
32 & 34.78 & 9,827 & 9,827 & 11,033 & 21,124 \\
33 & 35.99 & 10,488 & 10,488 & 11,792 & 21,729 \\
34 & 37.21 & 14,343 & 14,343 & 14,865 & 23,078 \\
35 & 38.41 & 18,106 & 18,106 & 17,805 & 23,500 \\
\bottomrule

		\end{tabular}
	\end{threeparttable}
	
						\vspace{4pt}
	\begin{minipage}{\textwidth}
		\footnotesize
		\textit{Notes:} The table reports heights (floors), estimated output quantities, the constrained maximum likelihood estimate (MLE), the MLE estimated separately by height, and the minimum and mean building prices (NIS per square meter).
	\end{minipage}
\end{table}

\clearpage
\FloatBarrier
	\subsection{Number of Observations by Height} \label{ap:tableobs}

\vspace{-0.2cm}
\begin{table}[!htb]
	\centering
	\begin{threeparttable}
		 \captionsetup{skip=4pt}
		\caption{Number of observations}
		\label{tablesumobs}
		\small
		\begin{tabular}{@{}r
				r
				c
				c
				r
				r@{}}
			\toprule
			Height 
			& Blocs 
			& \makecell{\% of blocs\\with one\\building}
			& \makecell{Mean \# of\\buildings\\per bloc}
			& Buildings 
			& Apartments \\
			\midrule
1 & 182 & 0.74 & 1.8 & 319 & 1,453 \\
2 & 629 & 0.53 & 2.6 & 1,662 & 8,101 \\
3 & 607 & 0.57 & 2.3 & 1,396 & 10,369 \\
4 & 874 & 0.45 & 3.4 & 2,968 & 28,266 \\
5 & 866 & 0.47 & 3.0 & 2,562 & 27,642 \\
6 & 826 & 0.49 & 2.8 & 2,315 & 27,336 \\
7 & 663 & 0.51 & 2.5 & 1,639 & 24,725 \\
8 & 572 & 0.53 & 2.3 & 1,340 & 24,086 \\
9 & 472 & 0.52 & 2.4 & 1,137 & 24,384 \\
10 & 341 & 0.55 & 2.0 & 675 & 15,741 \\
11 & 202 & 0.68 & 1.6 & 331 & 9,214 \\
12 & 155 & 0.64 & 1.6 & 253 & 7,517 \\
13 & 153 & 0.76 & 1.3 & 206 & 7,301 \\
14 & 121 & 0.69 & 1.7 & 202 & 6,369 \\
15 & 112 & 0.66 & 1.7 & 185 & 7,434 \\
16 & 93 & 0.68 & 1.5 & 142 & 6,024 \\
17 & 80 & 0.62 & 1.8 & 145 & 6,825 \\
18 & 75 & 0.71 & 1.6 & 121 & 4,041 \\
19 & 61 & 0.66 & 1.6 & 97 & 3,407 \\
20 & 62 & 0.73 & 1.5 & 90 & 3,744 \\
21 & 49 & 0.71 & 1.4 & 67 & 3,894 \\
22 & 42 & 0.69 & 1.6 & 69 & 2,373 \\
23 & 25 & 0.68 & 1.6 & 40 & 1,623 \\
24 & 36 & 0.78 & 1.2 & 45 & 1,930 \\
25 & 21 & 0.95 & 1.0 & 22 & 1,252 \\
26 & 18 & 0.78 & 1.4 & 26 & 902 \\
27 & 12 & 0.83 & 1.2 & 14 & 766 \\
28 & 15 & 0.67 & 1.4 & 21 & 925 \\
29 & 14 & 0.71 & 1.4 & 19 & 730 \\
30 & 14 & 0.86 & 1.1 & 16 & 659 \\
31 & 7 & 0.71 & 1.3 & 9 & 309 \\
32 & 7 & 1.00 & 1.0 & 7 & 205 \\
33 & 5 & 0.80 & 1.2 & 6 & 267 \\
34 & 6 & 0.83 & 1.3 & 8 & 267 \\
35 & 11 & 0.64 & 1.5 & 17 & 603 \\
\bottomrule

		\end{tabular}
	\end{threeparttable}
	
							\vspace{4pt}
	\begin{minipage}{\textwidth}
		\footnotesize
		\textit{Notes:}  The second, fifth, and sixth columns report the number of blocs, buildings, and apartments, respectively. The third column is the share of blocs that contain exactly one building of the given height. The fourth column is the mean number of buildings of that height per bloc. Conditional on height, the median number of buildings is one and the mean is about 2.4.
	\end{minipage}
\end{table}

\end{document}